\documentclass[twocolumn]{autart}

\usepackage{amsmath, amsfonts, mathtools, bm}				 
\usepackage{graphicx}
\usepackage{enumitem}
\usepackage{algorithm}
\usepackage{algpseudocode}
\graphicspath{{./fig/}}

\usepackage{standalone}
\usepackage{tikz}
\usetikzlibrary{calc, arrows.meta, positioning}

\newcommand{\SINR}{\mathrm{SINR}}
\newcommand{\Ptx}{P^\mathrm{tx}}
\newcommand{\Prc}{P^\mathrm{rc}}

\begin{document}

\begin{frontmatter}

\title{Transmission Power Allocation for Remote Estimation \\ with Multi-packet Reception Capabilities\thanksref{footnoteinfo}}
\thanks[footnoteinfo]{This paper was partially presented at the IFAC World Congress in Berlin 2020. Corresponding author: Matthias Pezzutto, mail: {\tt matthias.pezzutto@phd.unipd.it}.}

\author[unipd]{Matthias Pezzutto},
\author[unipd]{Luca Schenato},
\author[hamilton]{Subhrakanti Dey}

\address[unipd]{Department of Information Engineering, University of Padova, Padova, Italy}
\address[hamilton]{Hamilton Institute, National University of Ireland, Maynooth, County Kildare, Ireland}                      
                                         
\begin{abstract}
In this paper we consider the problem of transmission power allocation for remote estimation of a dynamical system in the case where the estimator is able to simultaneously receive packets from multiple interfering sensors, as it is possible e.g. with the latest wireless technologies such as 5G and WiFi. To this end we introduce a general model where packet arrival probabilities are determined based on the received Signal-to-Interference-and-Noise Ratio and with two different receivers design schemes, one implementing standard multi-packet reception technique and one implementing Successive Interference Cancellation decoding algorithm in addition. Then we cast the power allocation problem as an optimization task where the mean error covariance at the remote estimator is minimized, while penalizing the mean transmission power consumption. For the infinite-horizon problem we show the existence of a stationary optimal policy, while for the finite-horizon case we derive some structural properties under the special scenario where the overall system to be estimated can be seen as a set of independent subsystems. Numerical simulations illustrate the improvement given by the proposed receivers, especially when SIC is used, over orthogonal schemes that schedule only one sensor transmission at a time in order to avoid interference.
\end{abstract}


\end{frontmatter}

\section{Introduction}

Wireless networks and smart devices are quickly penetrating everyday life under the impulse of the Internet of Things, whose applications include smart homes, smart cities, wearable devices for advanced healthcare, and autonomous systems.
In many such applications, smart sensing devices monitor a dynamical system and communicate over wireless their measurements to a remote unit for state estimation and control input design.
In this context, a well-known problem is the transmission scheduling problem: for each time slot a subset of the sensors needs to be selected to transmit in order to optimize the remote estimation performance, while avoiding collisions or undesirable delays, and minimizing the transmission power consumption.
Based on the underlying dynamical nature of the source of data, many transmission scheduling strategies have been proposed in the literature: optimal periodic policies \cite{shi2011optimal}\cite{ren2013optimal}, general optimal (infinite-horizon) policies \cite{mo2014infinite}\cite{zhao2014optimal}, and event-trigger policies \cite{wu2012event}\cite{han2015stochastic}, possibly with packet loss \cite{leong2016sensor}.

In this paper, we consider the more general problem of power allocation for remote estimation, instead of the simpler transmission scheduling, that can be seen as the special case where only two power levels are allowed. Transmission power control on a packet basis is in fact allowed by the most recent wireless standards (e.g. see the case of IEEE 802.11ax \cite{afaqui2016ieee} and 5G \cite{olwal2016survey}) and supported by many off-the-shelf devices (a practical implementation on WiFi has been devised by \cite{huehn2012practical}).
To date, however, only a handful of works have considered this possibility. 
For instance, the case with a single sensor has been studied in \cite{han2013online}\cite{ren2014dynamic} with only two (non-null) power levels and in \cite{ren2017infinite} with a continuous bounded range of powers.

A key observation regarding the  aforementioned works is that they focus on the simplified case where interference is not present, either because only one sensor is available or because at most a sensor is scheduled for each time slot.
These solutions are indeed compliant with several existing wireless network standards whose medium access is regulated by Time-Division Multiple Access method, for instance the WirelessHART standard.
Following a different approach, simultaneous transmissions of multiple signals on the same time-frequency resources have recently been proposed for many network standards to increase the spectral efficiency.
Indeed, the capability of demodulating multiple signals in the presence of mutual interference, also known as multi-user detection, is well-known in the field of  broadband wireless communications (see \cite{verdu_MU_detect}).
Multi-packet reception is a particular type of multi-user detection technique where the receiver is equipped to decode multiple simultaneous transmissions. This can be achieved in many ways, such as at the signal modulation level (CDMA), by multiple antennas at transmitter and receiver (MIMO), or by using collision resolution methods based on signal processing as discussed in \cite{multi_packet_tong}.
These techniques have proven to be very appealing for next-generation wireless networks. In particular, 5G includes Non-Orthogonal Multiple Access (NOMA) that enables multi-packet reception using Successive Interference Cancellation (SIC). See \cite{islam2016power}\cite{ding2017survey} for a comprehensive overview, while a practical implementation is presented e.g. in \cite{xiong2015open}. NOMA together with SIC is also used in LTE-A \cite{lee2016multiuser} (in this context it is referred to as Multi-User Superposition Transmissions) and in latest digital TV standards \cite{zhang2016layered} (where it is referred to as Layered Division Multiplexing).
Besides its adoption in NOMA systems, SIC and other multi-packet reception techniques have been employed also in Wi-Fi \cite{ali2018uplink}, Long Range \cite{de2020lora}, and ZigBee \cite{kong2015mzig} networks.

Despite that multi-packet reception is supported by the latest wireless networks, this progress has not been widely explored for control applications so far.
To enable it, differently from existing aforementioned works on transmission scheduling and power allocation for remote estimation, mutual interference needs to be considered. A random access scheme, tailored for control applications, that does not exclude interference is proposed in \cite{gatsis2018random} but packets are assumed to be lost when interfering. 
Reception in presence of other interfering communications is considered in \cite{zoppi2019transmission} for the case of multiple independent systems with multiple estimators. Transmission powers are allocated in order to match minimum arrival probabilities for each system so that a minimum estimation quality is enforced.
A central estimator with multi-packet reception capabilities is explicitly considered only by a few works. The work \cite{li2014multi} addresses the power allocation problem in a game-theoretic framework and proposes a central coordinator to improve the performances.
The work \cite{li2019power} studies the properties of the one-step-ahead optimal power allocation policy for remote estimation when the transmission power can take continuous unbounded values.
Our previous work \cite{pezzutto2020transmission} considers a more general channel model than \cite{li2019power} and derives the finite-horizon optimal policy, but it is limited to the case with 2 power levels.

The aim of the paper is to introduce a tool to accommodate future sensor networks and control systems that will rely on next generation wireless networks supporting power selection at the transmitter and multi-packet reception at the receiver.
We introduce a very general system model including multiple sensors, multiple transmission power levels, a realistic fading channel model with mutual interference and arrival probabilities based on received Signal-to-Interference-and-Noise Ratio (SINR), and two different receiver design schemes, namely with and without SIC. Then, under the considered framework, an optimal power allocation strategy is determined by solving an (infinite or finite horizon) optimization problem that accounts for the average estimation quality and penalizes the total average transmission power.
The main contributions can be summarized as follows:
(i) for the first time we propose a transmission power allocation strategy that exploits SIC showing its benefits for remote estimation and not only for communication throughput maximization,
(ii) we prove the existence of an optimal stationary policy and the boundedness of the expected error covariance,
(iii) for the special case where the system actually consists of multiple independent subsystems, we show that the optimal one-step-ahead policy exhibits a threshold behaviour with respect to a scalar transformation of the error covariance and we characterize these thresholds for the case where only 2 power levels are available (namely the transmission is scheduled or not).
Interestingly, there always exists a set of error covariance matrices for which simultaneous transmissions at the maximum power is the optimal action. Even if this action entails a larger loss probability, it is optimal where error covariances of each subsystem are large enough.
The structural properties for the one-step-ahead policy simplify the implementation and give an interesting insight on the general optimal solution to derive possible heuristics.
Simulations on realistic systems show that SIC plays a key role to improve the estimation quality, allowing to halve the error covariance with respect to simpler coding-decoding algorithms in the case of a pendulum-on-a-cart system. These results motivate the application of SIC for remote estimation and, as a consequence, the employment of advanced wireless networks like 5G and WiFi.
The remainder of the paper is organized as follows: in Sec.~\ref{sec:problem} we introduce the main assumptions, in Sec.~\ref{sec:chanchar} we derive the characterization for the proposed channel model, in Sec.~\ref{sec:infpolicy} we introduce the optimal infinite-horizon power allocation problem and we show the existence and the stability of the optimal policy, in Sec.~\ref{sec:finpolicy} we study the optimal finite-horizon policy, while in Sec.~\ref{sec:decsys} we provide the structural properties for the case of independent subsystems. The paper ends with simulations in Sec.~\ref{sec:sim} and some concluding remarks.

\section{Problem formulation}\label{sec:problem}\vspace*{-5pt}

The outputs of a dynamical system are sampled by a set of $N$ sensors and communicated to a remote estimator through a wireless network. Each sensor is provided with a transmitter whose transmission power can be selected on a packet basis, while the estimator is equipped with a receiver with multi-packet reception capabilities so that it is able to simultaneously receive packets from multiple sensors. Further details are given in the following.

\subsection{System model}\label{sec:sysmodel}
Consider the discrete-time state-space linear model
\begin{equation}
x(k+1) = A x(k) + w(k)
\end{equation}
where $x(k) \in  \mathbb{R}^n$ is the state and $w(k)  \in  \mathbb{R}^n$ is the process noise modelled as independent and identically distributed (i.i.d.) Gaussian random variables $w(k) \sim \mathcal{N}(0,Q)$ with $Q\geq 0$. 
A set of $N$ sensors is available. At each time instant, the $i$-th sensor measures the output 
\begin{align}
y_i(k) = C_i x(k) + v_i(k)
\end{align}
where $y_i(k) \in  \mathbb{R}^{m_i}$ and $v_i(k)\in  \mathbb{R}^{m_i}$ is the measurement noise modelled as i.i.d. Gaussian random variables $v_i(k) \sim \mathcal{N}(0, \, R_i)$ with $R_i > 0$ and independent of $\{w(k)\}$.
During the $k$-th sampling period, a packet containing the sampled output $y_i(k)$ is communicated to the remote estimator using the transmission power $\Ptx_i(k)$. Due to path loss and fading caused by the wireless channel, the received power at the estimator $\Prc_i(k)$ is different from $\Ptx_i(k)$. 
The packet may be lost due to interference from other transmissions and channel and receiver noise. We represent this process through the variable $\gamma_i(k)$, which is equal to $1$ if the transmission of $y_i(k)$ is successfully completed, $0$ otherwise. 
Mathematically, we introduce a function $f_i:\mathbb{R}^N\rightarrow\{0,1\}$ to represent the underlying model of the arrival processes as
\begin{equation}
\gamma_i(k)=f_i(\Prc_1(k), \,\Prc_2(k), \cdots \,,\, \Prc_N(k)).
\end{equation}
The information set available to the central estimator at the time instant $k$ is:
\begin{align*}
\mathcal{I}(k) = \bigcup_{i=1}^{N} \mathcal{I}_i(k),  \ \
\mathcal{I}_i(k) =
\begin{multlined}[t]
\big\{\gamma_i(0), \gamma_i(0)y_i(0), \,\dots\, ,\\ 
\gamma_i(k\!-\!1), \gamma_i(k\!-\!1)y_i(k\!-\!1) \big\}
\end{multlined}
\end{align*}
where, with a little misuse of notation, if $\gamma_i(t) = 0$ then $\gamma_i(t)y_i(t)=$ \O, i.e. $y_i(t)$ is missing.
Define
\begin{align*}
\widehat{x}(k|k\!-\!1) \!&=\! \mathbb E[x(k) | \mathcal{I}(k)] \\
P(k|k\!-\!1) \!&= \!\mathbb{E}[(x(k) \!-\! \widehat{x}(k|k\!- \!1))(x(k) \!-\! \widehat{x}(k|k\!-\!1))' | \mathcal{I}(k)].
\end{align*}
From \cite{anderson2012optimal}, $\widehat{x}(k|k-1)$ is the optimal estimate of $x(k)$ based on the measurements available to the central unit up to time $k$, and the matrix $P(k|k-1)$ is the corresponding estimation error covariance matrix.
In general, how to obtain the closed forms of $\widehat{x}(k|k-1)$ and $P(k|k-1)$ is a challenging problem. In the considered setting with intermittent observations from multiple sensors, they have been obtained in \cite{garone2011lqg}.
In order to more easily manage the update of the error covariance matrix, we arrange the results of \cite{garone2011lqg} with the information form of the optimal estimator given by \cite{hashemipour1988decentralized}, obtaining
\begin{align}
&P(k|k) = \Bigg(P^{-1}(k|k-1)  + \sum\limits_{i=1}^{N} \gamma_i(k) C'_iR^{-1}_iC_i \Bigg)^{-1} \\
&P(k+1|k) = A P(k|k) A' + Q
\end{align}
starting from $P(0)=P(0|0)=\mathbb{E}[x(0)x(0)']$ and
\begin{align*}
&\begin{multlined}[t]
\hat{x}(k+1|k) = \\ 
A\hat{x}(k|k-1)  + AK(k)\Gamma(k)(y(k) + C\hat{x}(k|k-1))
\end{multlined}\\
&\begin{multlined}[t]
K(k) = \\ 
P(k|k-1) C'\Gamma(k)'(\Gamma(k)(CP(k|k-1)C' + R)\Gamma(k)' )^{-1} 
\end{multlined}
\end{align*}
where $\Gamma(k)=\textrm{diag}\{\gamma_i(k)\mathrm{I}_{m_{i}}\}$ and $\mathrm{I}_{m_{i}}$ is the identity matrix of dimension $m_i$.
Equations above are obtained by applying the standard derivation of the Kalman filter for the time-varying system with output matrix $\Gamma(k)C$ and using the Matrix Inversion Lemma on the error covariance update.
The transmission powers $\Ptx_i(k)$,  $i=1, \dots, N$ are chosen at the estimator side and are sent back to the sensors within the time interval $(k-1,k)$.
\begin{rem}
	The remote estimator is assumed to be provided with an unlimited energy budget (e.g. when it is connected to the main power supply such as in a central control room) so that its transmission power can be assumed high enough for $\Ptx_i(k)$ to be communicated without error. 
	Note that the transmission powers can be encoded using a few bits (as in the case where the number of sensors and powers is relatively small) so that redundancy can	be used in coding to achieve high reliability without wasting bandwidth.
\end{rem}

\begin{figure}
	\centering
	\begin{tikzpicture}	 

	\coordinate (sys) at (0,0);
	\coordinate (sens1) at (2,0.5);
	\coordinate (tx1) at (3.7,0.5);
	\coordinate (sens2) at (2,0);
	\coordinate (tx2) at (3.7,0);
	\coordinate (txN) at (3.7,-0.5);
	\coordinate (sensN) at (2,-0.5);
	\coordinate (est) at (6.7,0);
	
	\node[draw, minimum width=1.7cm, minimum height=1.5cm, line width=1pt, align=center] (SYS) at (sys) {System};
	\node[draw, minimum width=1.7cm, line width=1pt, align=center] (SENS1) at (sens1) {Sensor 1};
	\node[minimum width=1.7cm, line width=1pt, align=center] (SENS2) at (sens2) {\dots};
	\node[draw, minimum width=1.7cm, line width=1pt, align=center] (SENSN) at (sensN) {Sensor $N$};
	\node[draw, minimum width=0.7cm, line width=1pt, align=center] (TX1) at (tx1) {TX};
	\node[draw, minimum width=0.7cm, line width=1pt, align=center] (TXN) at (txN) {TX};
	\node[draw, minimum width=1.7cm, minimum height=1.5cm, line width=1pt, align=center] (EST) at (est) {Estimator};
	
	\coordinate (switch3-START) at ($(TX1)+(0.9,0)$);
	\coordinate (switch3-CLOSED) at ($(switch3-START)+(0.5,0)$);
	\coordinate (switch3-OPEN) at ($(switch3-START)+(0.4325,0.25)$);	
	\coordinate (switch4-START) at (switch3-START |- TXN);
	\coordinate (switch4-CLOSED) at ($(switch4-START)+(0.5,0)$);
	\coordinate (switch4-OPEN) at ($(switch4-START)+(0.4325,0.25)$);	
	
	\draw[-latex] (SYS.east |- SENS1.west) -- (SENS1.west) node[left, pos=0.5]{};
	\draw[-latex] (SYS.east |- SENSN.west) -- (SENSN.west) node[left, pos=0.5]{ };
	\draw[-latex] (SENS1) -- (TX1) node[pos=0.5, above]{$y_1$}; 
	\draw[-latex] (SENSN) -- (TXN) node[pos=0.5, above]{$y_{N}$}; 
	\draw[-] (TX1) -- (switch3-START) node[pos=0.5, above]{}; 
	\draw[-] (switch3-START) -- (switch3-OPEN); 
	\draw[-latex] (switch3-CLOSED) --  (EST.west |- switch3-CLOSED); 
	\draw[-] (TXN) -- (switch4-START) node[pos=0.5, above]{}; 
	\draw[-] (switch4-START) -- (switch4-OPEN); 
	\draw[-latex] (switch4-CLOSED) --  (EST.west |- switch4-CLOSED); 
	
	\draw[->, line width=1pt] ($(switch3-START)+(0,0.25)$) to [in=90] ($(switch3-CLOSED)-(0.25,0.1)$);
	\draw[->, line width=1pt] ($(switch4-START)+(0,0.25)$) to [in=90] ($(switch4-CLOSED)-(0.25,0.1)$);
	\node[] (l) at ($(switch3-START)+(0.25,0.4)$) {$\gamma_1$};
	\node[] (l) at ($(switch4-START)+(0.25,-0.3)$) {$\gamma_N$};
	\node[] (l) at ($(SYS)+(4.9,0)$) {\dots};
	
	\pgfmathsetmacro{\x}{5.5}
	\pgfmathsetmacro{\y}{-0.6}
	\pgfmathsetmacro{\s}{0.6}
	\pgfmathsetmacro{\l}{1.2}
	\begin{scope}[shift={(5.5,5.25)},rotate=270]
	\draw (-1.6*\s+\x,-0.7*\s*\l+\y) .. controls (-2.3*\s+\x,-1.1*\s*\l+\y)
	and (-2.7*\s+\x,0.3*\s+\y) .. (-1.7*\s+\x,0.3*\s+\y) .. controls (-1.6*\s+\x,0.7*\s+\y)
	and (-1.2*\s+\x,0.9*\s+\y) .. (-0.8*\s+\x,0.7*\s+\y) .. controls (-0.5*\s+\x,1.5*\s+\y)
	and (0.6*\s+\x,1.3*\s+\y) .. (0.7*\s+\x,0.5*\s+\y) .. controls (1.5*\s+\x,0.4*\s+\y)
	and (1.2*\s+\x,-1*\s*\l+\y) .. (0.4*\s+\x,-0.6*\s*\l+\y) .. controls (0.2*\s+\x,-1*\s*\l+\y)
	and (-0.2*\s+\x,-1*\s*\l+\y) .. (-0.5*\s+\x,-0.7*\s*\l+\y) .. controls (-0.9*\s+\x,-1*\s*\l+\y)
	and (-1.3*\s+\x,-1*\s*\l+\y) .. cycle;
	\end{scope}	
\end{tikzpicture}\vspace{-5pt}
	\caption{Setup. Transmitters (TX) can choose among a finite set of possible powers.}
	\label{fig:setup}
\end{figure}
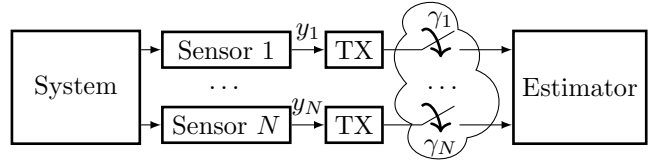

\subsection{Channel model}
The wireless medium is modelled as a fading channel with Additive White Gaussian Noise (AWGN) whose average power at the remote estimator is $\sigma^2$. 
The transmission power $\Ptx_i(k)$ of the $i$-th sensor is selected from a finite set $\mathcal{P}_i$
\begin{equation}
\mathcal{P}_i = \left\{0,\, \Ptx_{i,1},\, \Ptx_{i,2},\, \dots\,,\, \Ptx_{i,\text{max}} \right\}.
\end{equation}
Let $s_i$ denote the slow fading component of the channel power gain (usually dependent on path loss etc.) from the $i$-th sensor to the remote estimator, while $r_i(k)$ is the fast fading component of the same channel during the $k$-th sampling period. 
We assume that $s_i$ is constant, while $r_i(k)$ is modelled as a temporally i.i.d. exponential random variable
with unity mean, i.e. $r_i(k)\sim Exp(1)$, and we also assume that $r_i(k),  r_j(t)$ are mutually independent for $\forall k,t$ and $i\neq j$.
Then, the received power (at the remote estimator) $\Prc_i(k)$ from the $i$-th sensor is
\begin{align}\label{eq:rcpow}
\Prc_i(k) = s_i  r_i(k) \Ptx_i(k).
\end{align}
It follows that the received power $\Prc_i(k)$ is an exponential random variable with mean $\lambda_i(k) = (s_i\Ptx_i(k))^{-1}$, i.e. $\Prc_i(k) \sim Exp(\lambda_i(k))$. Due to the independence of fast fading gains $r_i(k)$, received powers are temporally and spatially independent, namely $\Prc_i(k)$ is independent of $\Prc_i(t) \ t \neq k$ and of $\Prc_j(t) \ \forall k,t$ and $j\neq i$.
This model, commonly referred to as Rayleigh fading, relates to the case when the number of reflected paths is large, such as in a factory or in an urban environment with many scatterers. In that case, the received electric field can be decomposed into two orthogonal components: each one is given by the sum of a large number of (zero-mean) terms and so it can be approximated by a Gaussian random variable. Therefore, the channel power gain is the sum of the squares of two Gaussian random variables, resulting in an exponential distribution (see e.g. \cite[Ch. 3.2]{goldsmith2005wireless} and \cite[Ch. 14.1]{proakis2001digital}).
In the following we denote $\Prc(k)=(\Prc_1(k), \,\Prc_2(k),\, \dots \,,\, \Prc_N(k))$ and $\Ptx(k)=(\Ptx_1(k), \,\Ptx_2(k),\, \dots \,,\, \Ptx_N(k))$.

\subsection{Receiver model}
In contrast to most of the literature on transmission scheduling for remote estimation, we assume that the receiver has multi-packet reception capabilities, namely, it is able to decode multiple incoming signals. 
Successful reception of a packet requires that the signal carrying the information is strong enough with respect to the amount of interference and noise that disturb the transmission. The required strength depends on communication parameters like the modulation, the coding scheme, and the symbol rate.
In this view, following e.g. \cite{ephremides2002energy}, the arrival process of the packet containing $y_i(k)$ is modeled
\begin{equation}\label{eq:arrvar}
\gamma_i(k) = f_i(\Prc(k))=
\begin{cases}
1 & \text{if } \, \SINR_i(\Prc(k)) > \alpha \\
0 & \text{otherwise}
\end{cases}
\end{equation}
where $\SINR_i(\Prc(k))$ is the Signal-to-Interference-and-Noise Ratio (at the decoding instant) corresponding to the packet containing $y_i(k)$, while $\alpha > 0$ is the reception threshold set by the communication parameters.
Since the received powers are independent across time slots, $\gamma_i(k)$ is an i.i.d. Bernoulli process. However  $\gamma_i(k), \gamma_j(k)$ for $j \neq i$ may be dependent on each other due to interference within a given time slot. In this paper, we consider two different receiver design schemes, that result in two different expressions for the SINR.

\subsubsection{Simple receiver}
The first receiver, referred to as simple receiver, decodes each signal independently so that the SINR is
\begin{align}\label{eq:sinr:simple}
\SINR_i(\Prc(k)) 
= \frac{\Prc_i(k)}{\sum\limits_{j \neq i}^{} \Prc_j(k)+ \sigma^2}.
\end{align}
According to this definition, the transmission from sensor $i$ is affected by the interference of the transmissions from each other sensor $j\neq i$ \cite[Ch. 6.1.1]{goldsmith2005wireless}\cite{jindal2004duality}. In typical information theoretic fashion, the received interference power is simply added to the noise power. The underlying  assumption is that interference is statistically independent of combined channel and receiver noise. 
This is motivated by the use of random Gaussian codebooks at the transmitters to achieve channel capacity for various types of fading channels, such as multiple-access, broadcast, and interference channels \cite{jindal2004duality}.
In this case we need to have $\alpha \in (0,\, 1)$ to enable multi-packet reception.

\subsubsection{Successive Interference Cancellation}
The second examined receiver differs from the first for the implementation of Successive Interference Cancellation (SIC) algorithm.
In this case, the sensor with the highest received power is decoded first and the reconstructed signal is subtracted out from the total received signal so that the second strongest signal can be decoded next when the strongest interference is no more present.
In this way, after the first successful reception, other packets are decoded under an improved SINR and the probability of simultaneous received packets is definitely increased. 
According to formal information theoretic treatment \cite{tse_ITbook}, sensors are usually relabelled by sorting the received powers in descending order so that the interference power is due only to subsequent sensors in the new order.
We equivalently define the set $\mathcal{J}_i(k)$ containing the indices of the sensors whose received powers are higher than the received power from $i$-th sensor
\begin{equation}
\mathcal{J}_i(k) = \left\{j \,:\, \Prc_j(k) > \Prc_i(k) \right\}.
\end{equation}
Then, the SINR for the $i$-th sensor in this case is given by
\begin{multline}\label{eq:sinr:sic}
\SINR_i(\Prc(k))\\
=\begin{cases}
\dfrac{\Prc_i(k)}{\sum\limits_{\substack{j\notin\mathcal{J}_i(k) \\ j\neq i}} \Prc_j(k) + \sigma^2} &
\begin{multlined}[c]
\text{if } \SINR_j(\Prc(k))>\alpha \vspace*{-5pt} \\ \text{for any } j \in \mathcal{J}_i(k)
\end{multlined}
\vspace*{5pt}\\
\dfrac{\Prc_i(k)}{\sum\limits_{j\neq i}^{} \Prc_j(k)+ \sigma^2}  &\text{otherwise}.
\end{cases}
\end{multline}
According to this definition, the transmission from sensor $i$ is affected by the interference of the transmissions from each other sensor not decoded yet, possibly all the sensors $j \neq i$ if packets with higher received power have not been decoded \cite{tse_ITbook}.
In this case, even if $\alpha \geq 1$, multi-packet reception is possible.

\begin{rem}
The above SINR model provides an accurate approximation of the arrival processes when the interfering transmissions are synchronized or overlap for a considerable amount of time. This can be achieved through internal synchronization procedures in the Physical layer of the OSI stack, so it can be assumed without loss in terms of applicability. 
In line with existing works, also sampling instants of the multiple sensors are assumed to be synchronized. 
Nonetheless, this assumption can be relaxed if the temporal displacement between sampling at different sensors is small enough to be negligible with respect to the sampling period.
\end{rem}

\section{Channel characterization}\label{sec:chanchar}

\begin{figure}[t]
	\begin{minipage}{0.45\linewidth}
		\centering
		\begin{tikzpicture}
	\coordinate (x0) at (-0.5,0);
	\coordinate (y0) at (0,-0.5);
	\coordinate (x) at (3,0);
	\coordinate (y) at (0,3);
	
	\coordinate (r1start) at (1,0);
	\coordinate (r1end) at (2.2,3);
	\coordinate (r2start) at (0,1);
	\coordinate (r2end) at (3,2.2);
	\coordinate (rmid) at (1.7,1.7);
	\coordinate (rmidbott) at (1.7,0);
	\coordinate (rmidleft) at (0,1.7);

	\draw[-{Latex[length=2mm, width=1.5mm]}, thin] (x0) --  (x) node[pos=0.9, below] {\footnotesize $P^\mathrm{rc}_1$};
	\draw[-{Latex[length=2mm, width=1.5mm]}, thin] (y0) -- (y) node[pos=0.9, left] {\footnotesize $P^\mathrm{rc}_2$};
	
	\draw[-, line width=0.5mm] (r1start) -- (r1end) node[pos=0, yshift=-4pt] {\tiny $ \alpha\sigma^2$} ;
	\draw[-, line width=0.5mm] (r2start) -- (r2end) node[pos=0, xshift=-8pt] {\tiny $ \alpha\sigma^2$};
	\draw[dashed, line width=0.1mm] (rmidbott) -- (rmid) node[pos=0, yshift=-6.5pt] {\tiny $ \frac{\alpha\sigma^2}{1\!-\!\alpha}$} ;
	\draw[dashed, line width=0.1mm] (rmidleft) -- (rmid) node[pos=0, xshift=-8pt] {\tiny $ \frac{\alpha\sigma^2}{1\!-\alpha}$} ;
	
	\node[](l) at (0.55,0.5){$R(0,0)$};
	\node[](l) at (0.8,2.2){$R(0,1)$};
	\node[](l) at (2.4,0.8){$R(1,0)$};
	\node[](l) at (2.6,2.45){$R(1,1)$};
	
\end{tikzpicture}
	\end{minipage}\hfill
	\begin{minipage}{0.45\linewidth}
		\centering
		\begin{tikzpicture}
	\coordinate (x0) at (-0.5,0);
	\coordinate (y0) at (0,-0.5);
	\coordinate (x) at (3,0);
	\coordinate (y) at (0,3);
	
	\coordinate (r1start) at (1.1,0);
	\coordinate (r1end) at (1.7,1.7);
	\coordinate (r2start) at (0,1);
	\coordinate (r2end) at (1.7,1.7);
	\coordinate (r3start) at (1.1,24/17);
	\coordinate (r3end) at (1.1,3);
	\coordinate (r4start) at (24/17,1);
	\coordinate (r4end) at (3,1);
	\coordinate (rmidbott) at (1.7,0);
	\coordinate (rmidleft) at (0,1.7);

	\draw[-{Latex[length=2mm, width=1.5mm]}, thin] (x0) --  (x) node[pos=0.9, below] {\footnotesize $P^\mathrm{rc}_1$};
	\draw[-{Latex[length=2mm, width=1.5mm]}, thin] (y0) -- (y) node[pos=0.9, left] {\footnotesize $P^\mathrm{rc}_2$};
		
	\draw[-, line width=0.5mm] (r1start) -- (r1end);
	\draw[-, line width=0.5mm] (r2start) -- (r2end);
	\draw[-, line width=0.5mm] (r3start) -- (r3end);
	\draw[-, line width=0.5mm] (r4start) -- (r4end);
	\draw[dashed, line width=0.1mm] (r1start) -- (r3start) node[pos=0, yshift=-6.5pt] {\tiny $ \alpha\sigma^2$} ;
	\draw[dashed, line width=0.1mm] (r2start) -- (r4start)node[pos=0, xshift=-8pt] {\tiny $ \alpha\sigma^2$} ;
	
	\draw[dashed, line width=0.1mm] (rmidbott) -- (r1end) node[pos=0, yshift=-6.5pt] {\tiny $ \frac{\alpha\sigma^2}{1\!-\!\alpha}$} ;
	\draw[dashed, line width=0.1mm] (rmidleft) -- (r2end) node[pos=0, xshift=-8pt] {\tiny $ \frac{\alpha\sigma^2}{1\!-\alpha}$} ;

	\node[](l) at (0.5,0.5){$R(0,0)$};
	\node[](l) at (0.5,2.2){$R(0,1)$};
	\node[](l) at (2.4,0.5){$R(1,0)$};
	\node[](l) at (2.5,2.5){$R(1,1)$};
	
\end{tikzpicture}
	\end{minipage}\vspace{-5pt}
	\caption{Possible received powers and corresponding outcome of the arrival process. Left: without SIC. Right: with SIC.}
	\label{fig:arrivalregion}
\end{figure}
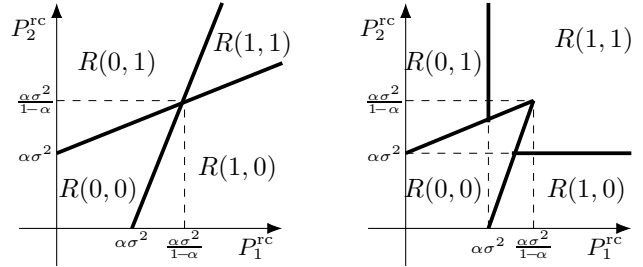

In this section we provide the probabilities of the arrival process as a function of the transmission powers. To this end, if $\Ptx_i>0$ then $\lambda_i=(\Ptx_is_i)^{-1}$ is well-defined and the joint distribution of received powers given the transmitted ones is
\begin{equation}
p(\Prc_1,\dots,\Prc_N\,|\,\Ptx_1,\dots,\Ptx_N) = \prod_{i=1}^{N}\lambda_i e^{\lambda_i\Prc_i}
\end{equation}
since channel gains are independent. Adapted to the  $\sigma$-algebra of the random variable $\Prc$, we can define the subset $R(\gamma)$ for $\gamma=(\gamma_1,\dots, \gamma_N)$ as
\begin{equation}
R(\gamma) = \left\{\Prc \,:\,  f_i(\Prc)=\gamma_i \ \ i=1,\,\dots\,,\,N\right\}
\end{equation}
that is the subset of the possible received powers indicating whether packets  have arrived or not as defined by $\gamma$. Explicit expressions for sets $R(\gamma)$ can be retrieved by combining \eqref{eq:arrvar} with \eqref{eq:sinr:simple} or \eqref{eq:sinr:sic}.
Note that they are different in general for the case with and without SIC, see for example the case with two sensors reported in Fig. \ref{fig:arrivalregion}. A remarkable exception is $R(0,\dots,0)$, namely where no packets are correctly received, since, if the packet with the highest received power is not decoded, no advantages are given by SIC.
Finally we can obtain the probability $P(\gamma|\Ptx=u)$ of $\gamma$ given the transmitted power $\Ptx\!=\!u$ as the $N$-dimensional integral
\begin{equation}\label{eq:def:prob}
P(\gamma\,|\,\Ptx\!=\!u) = \int \!\!\! \int \!\dots\! \int_{R(\gamma)}p(\Prc\,|\,\Ptx\!=\!u)d\Prc
\end{equation}
The arrival probabilities can be computed either analytically or numerically.
In the case where $\Ptx_i=0$, $\lambda_i$ is not defined and the joint distribution needs to be modified accordingly in order to consider that $\Prc_i=0$ with probability 1. The expression is omitted for sake of readability.

\begin{rem}
Spatial independence among fading channels from different sensors is quite a common assumption in wireless communications. Indeed, with high carrier frequencies, it can be ensured by a (small) sufficient separation between antennas at the transmitter/receiver (in the range of several millimeters for millimeter-wave communications in 5G for example).
While spatial independence allows easier calculations of the arrival packet probabilities, spatially correlated fading channels can be also considered at the expense of more complex analytical computation.
Similarly, temporally correlated fading channels can also be considered via models such as Markov fading channels. In that case, the power allocation problem considered in this paper can be solved by considering an augmented Markov chain including the channel process in the Markov Decision Process presented in next sections.
\end{rem}

In the following we highlight the dependence of arrival random variables $\gamma_i$ on the transmitted powers $\Ptx=u$ through \eqref{eq:rcpow} and \eqref{eq:arrvar} by writing $\gamma_i(u)$, and its expected value is denoted by
\begin{equation}
p_i(u)=P(\gamma_i=1\,|\,\Ptx=u)=\mathbb{E}[\gamma_i(u)].
\end{equation}
It is easy to show that the probability $p_i(u)$ of receiving a packet from sensor $i$ is monotonically increasing with respect to the power $\Ptx_i$ allocated to sensor $i$ and monotonically decreasing with respect to $\Ptx_j$ for $j\neq i$.

We conclude this section providing the exact expressions for the case with 2 sensors, as an example.
We denote by $p_{ij}(u)$ the probability that $\gamma_1 = i$ and $\gamma_2 = j$, with $i,j\in\{0,1\}$, under the allocated power $u=([u]_1,[u]_2)$. More specifically,
\begin{align*}
p_{11}(u) = \bm{P} (\gamma_1 = 1, \, \gamma_2 = 1 \, | \, \Ptx_1\!=\![u]_1, \,  \Ptx_2\!=\![u]_2) \\
p_{10}(u) = \bm{P} (\gamma_1 = 1, \, \gamma_2 = 0 \, | \, \Ptx_1\!=\![u]_1, \,  \Ptx_2\!=\![u]_2) \\
p_{01}(u) = \bm{P} (\gamma_1 = 0, \, \gamma_2 = 1 \, | \, \Ptx_1\!=\![u]_1, \,  \Ptx_2\!=\![u]_2) \\
p_{00}(u) = \bm{P} (\gamma_1 = 0, \, \gamma_2 = 0 \, | \, \Ptx_1\!=\![u]_1, \,  \Ptx_2\!=\![u]_2)
\end{align*}
If $\Ptx_1>0$ and $\Ptx_2>0$, under the assumption that $\alpha \in (0,1)$, we have
\begin{align*}
p_{11} 
&=\left( \frac{\lambda_1}{\lambda_1 + \alpha \lambda_2} +  \frac{\lambda_2}{\lambda_2 + \alpha \lambda_1} - 1 \right) e^{-(\lambda_1 + \lambda_2)\frac{\alpha}{1-\alpha}\sigma^2} \\
p_{10}&=\frac{\lambda_2}{\lambda_2 + \alpha \lambda_1} \, e^{-\alpha \lambda_1 \sigma^2} -p_{11}\\
p_{01}&=\frac{\lambda_1}{\lambda_1 + \alpha \lambda_2} \, e^{-\alpha \lambda_2 \sigma^2} - p_{11}\\
p_{00} 
&= 1 - (p_{11} + p_{10}) - (p_{11} + p_{01}) + p_{11} \\
&=\begin{multlined}[t]
1  - \frac{\lambda_2}{\lambda_2 + \alpha \lambda_1} \, e^{-\alpha \lambda_1 \sigma^2} - \frac{\lambda_1}{\lambda_1 + \alpha \lambda_2} \, e^{-\alpha \lambda_2 \sigma^2} + \\
\left( \frac{\lambda_1}{\lambda_1 + \alpha \lambda_2} +  \frac{\lambda_2}{\lambda_2 + \alpha \lambda_1} - 1 \right) e^{-(\lambda_1 + \lambda_2)\frac{\alpha}{1 - \alpha}\sigma^2} \vphantom{\Bigg(}
\end{multlined}
\end{align*}
If $\Ptx_1>0$ and $\Ptx_2=0$ we have
\begin{equation*}
p_{11}=0 \quad p_{10}=e^{-\alpha\lambda_1\sigma^2} \quad p_{01}=0 \quad p_{00}=1-e^{-\alpha\lambda_1\sigma^2}
\end{equation*}
and similarly for $\Ptx_1=0$ and $\Ptx_2>0$.
Clearly, if $\Ptx_1=\Ptx_2=0$ then $p_{11}=p_{10}=p_{01}=0$, $p_{00}=1$.

When SIC is employed, the corresponding probabilities (denoted by the superscript SIC) are given by 
\begin{align*}
p_{11}^{SIC} \! &= \!
\left(\! \frac{\lambda_2}{\lambda_2 \!+\! \alpha \lambda_1} e^{-\lambda_1 \alpha \sigma^2}
\!+\! \frac{\lambda_1}{\lambda_1 \!+\! \alpha \lambda_2} e^{-\lambda_2 \alpha \sigma^2} \! \right) \! e^{-\left(\lambda_1 + \lambda_2 \right) \alpha \sigma^2}  \\
& \quad \quad  \ + \left(1 - \frac{\lambda_2}{\lambda_2+ \alpha \lambda_1} - \frac{\lambda_1}{\lambda_1+ \alpha \lambda_2} \right)  e^{-(\lambda_1+\lambda_2)\frac{\alpha}{1-\alpha}\sigma^2} \\
p_{10}^{SIC} &= \frac{\lambda_2}{\lambda_2+ \alpha \lambda_1} e^{-\alpha \sigma^2 \lambda_1} \left(1 - e^{-\alpha \sigma^2\left(\lambda_2 + \alpha\lambda_1\right)}\right) \\
p_{01}^{SIC} &= \frac{\lambda_1}{\lambda_1+ \alpha \lambda_2} e^{-\alpha \sigma^2 \lambda_2} \left(1 - e^{-\alpha \sigma^2\left(\lambda_1 + \alpha\lambda_2\right)}\right)\\
p_{00}^{SIC} &= p_{00}.
\end{align*}
The cases where at least one transmitted power is null are equivalent to the case without SIC.
It can be shown that arrival probabilities $p_i(u)$ are increased by SIC. This can be understood also by comparing left and right panels of Fig.~\ref{fig:arrivalregion}, where the larger area of region $R(1,1)$ indicates the improved decoding capabilities of SIC.

\section{Optimal power allocation: infinite horizon}\label{sec:infpolicy}

We cast the power allocation task as a stochastic control problem. In particular, we define the state space $\mathcal{X}$ as the set of positive definite matrices of dimension $n \times n$, and the action space $\mathcal{U} \subset \mathbb{R}^N$ as the set of possible combinations of transmission powers, namely 
\begin{equation}
\mathcal{U} = \left\{ \Ptx=(\Ptx_1, \,\dots\, ,\,\Ptx_N) : \Ptx_i \in \mathcal{P}_i \right\}
\end{equation}
We are interested in an infinite-horizon policy 
\begin{equation}
U=\left\{u_1, u_2, u_3 \dots \right\}
\end{equation}
where $u_k : \mathcal{X} \rightarrow \mathcal{U}$ is the function mapping the states $P(k|k-1)$ into the control action $\Ptx(k) \!=\! u_k(P(k|k\!-\!1))$, that minimizes the infinite-horizon discounted cost
\begin{align}
J&\left(U, P(0)\right)= \nonumber \\
&\lim\limits_{K \rightarrow \infty} \mathbb{E}\left[\sum_{k=0}^{K-1} \beta^k\mathcal{C}(P(k\!+\!1|k),u_{k+1}(P(k\!+\!1|k)))\Bigg|P(0)\right]
\end{align}
where $\beta\in(0,1)$ is the discount factor and $\mathcal{C}(P,u)$ is one-step-ahead cost from state $P$ with action $u$. 
The use of a discount factor is common in infinite-horizon stochastic control problems to put more emphasis on current cost terms and less importance on
costs incurred in the distant future. The discounted problem is also more amenable from the theoretical point of view since the assumptions for the existence of the stationary optimal policy are milder than those for the average cost problem. However, the latter can still be considered at the expense of added technical assumptions.

In this work we choose
\begin{equation}\label{eq:onestepcost}
\mathcal{C}(P,u) = \mathbb{E}[\mathrm{Tr}(g(P,u))|P,u] +  \bar{\mu}u
\end{equation}
where $\bar{\mu}= (\mu, \, \mu, \,\dots\, , \,\mu)$ is a regularization parameter and the function $g(P,u)$ is the Riccati-like operator
\begin{equation}
g(P,u) = A\left(P^{-1} \!+\! \sum_{i=1}^{N}\gamma_i(u)C'_iR_i^{-1}C_i\right)^{\!-1}\!A' + Q.
\end{equation}
The cost balances the estimation quality, given by the trace of the error covariance, and the transmission power consumption of the sensors through the regularization parameter $\mu$.  
Minimizing the cost for different values of $\mu$ corresponds to minimizing the error covariance under different energy constraints: a larger $\mu$ implies a more stringent energy budget.
Recall that transmissions from the central unit are not penalized, since it is assumed to have an unlimited energy budget (see also Remark 1).
Formally the problem is then
\begin{equation}\label{eq:problem}
U^*=\arg\min_{U} J\left(U, P(0)\right)
\end{equation}
resembling a discounted infinite-horizon Markov Decision Process (MDP). 
The fact that the resulting stochastic control problem can be formulated via an MDP follows from the definition of MDP with state space given by the set of error covariance matrices, the action space given by the possible transmission powers, the above cost function definitions, and the fact that the state transition probability only depends on the previous state (error covariance) and the current control action (allocated power). This is due to the Markov property of the error covariance update. The proof is not detailed here for reasons of space.

In general, the problem may not admit a solution. Moreover, even if the solution exists, the optimal action for a given error covariance may be time-dependent and thus not tractable for  practical implementation. Several results are available in the literature of MDP, providing sufficient conditions for the existence of an optimal solution that is stationary, namely that does not depend on time. A very strong condition is to have the one-step-ahead cost $\mathcal{C}(P,u)$ bounded, that is not true in our case. A weaker sufficient condition requires the existence of a positive scalar $m >0$ and of a measurable function $w:\mathcal{X}\rightarrow\mathbb{R}^+$ such that
\begin{equation*}
\mathcal{C}(P,u)<mw(P) \quad \text{ and } \quad \sum_{z\in \mathcal{X}} w(z)\bm{P}(z|P,u) <  w(P)
\end{equation*}
for any pair $(P,u)$, $P\in\mathcal{X}$, $u\in\mathcal{U}$. Roughly speaking, this requires the cost to be (bounded by a function that is) a contraction in mean for any possible action. 
The most general condition to prove is the existence of a policy $\bar{U}=(\bar{u}_1,\bar{u}_2,\dots)$ such that $J(\bar{U},P(0))<\infty$ for any $P(0)\in \mathcal{X}$. See \cite{hernandez1992discrete} for a full treatment.
In the remainder of this section, based on the last condition, we show that a stationary optimal policy exists if it is possible to keep the evolution of the expected error covariance bounded.

Let $\mathcal{J}$ be a set of indices of sensors. Let the output matrix $C_\mathcal{J}$ denote the stacked version of the output matrices $C_i$ for $i\in\mathcal{J}$.
Among all the possible sets, choose a set $\mathcal{J}$ such that $(A,C_\mathcal{J})$ is detectable. 
Denote by $R_\mathcal{J}$ the corresponding block-diagonal matrix obtained from the matrices $R_i$ for $i\in\mathcal{J}$. 
Then, for the given set, we can introduce two power allocation policies,  characterized, respectively, by the perfect multi-packet reception probability, defined as
\begin{align}
p_\mathrm{mp} \!=\! \bm{P}(\gamma_i\!=\!1,\, i\!\in\!\mathcal{J}\,|\, \Ptx_i\!=\!\Ptx_{i,\mathrm{max}},\,i\!\in\!\mathcal{J}, \, \Ptx_j\!=\!0,\,j\!\notin\!\mathcal{J} )
\end{align}
and the worst-channel arrival probability, defined as
\begin{align}
p_\mathrm{wc} = \min_i \bm{P}(\gamma_i\!=\!1\,|\, \Ptx_i\!=\!\Ptx_{i,\mathrm{max}},\, \Ptx_j\!=\!0, \, j\!\neq\!i )
\end{align}
Let $\Lambda(A)=1-1/\prod_\ell |\lambda_\ell^u(A)|^2$ where $\lambda_\ell^u(A)$ is the $\ell$-th unstable eigenvalue of $A$.
Then we have the following sufficient stability condition.
\begin{lem}\label{th:stability}	Let the set $\mathcal{J}$ be such that the pair $(A,C_\mathcal{J})$ is detectable and let the pair $(A,\sqrt{Q})$ be reachable.
Assume that at least one of the following conditions holds
\begin{enumerate}[topsep=-10pt]
	\item  $p_\mathrm{mp}> \Lambda(A)$
	\item  $p_\mathrm{wc}> \Lambda(A^{|\mathcal{J}|})$
\end{enumerate}
Then there exists a policy $\bar{U}$ such that $\exists M_{P(0)}>0$ for which $\mathbb{E}[P(k|k-1)]\leq M_{P(0)}$ for any $k>0$.
\end{lem}\vspace{-15pt}
\begin{pf} See Appendix A. \end{pf}\vspace{-15pt}
The previous lemma provides two sufficient but not necessary conditions for the boundedness of the error covariance, equivalent to the mean-square stability of the estimator.
The first one is related to the arrival rate when all the measurements required for the system to be detectable are simultaneously transmitted. 
The second one relates the stability of the remote estimate to the characteristic of the worst channel and to the under-sampled system, as if all the measurements are transmitted in a unique packet every $|\mathcal{J}|$ sampling periods.
Note that the set of indices $\mathcal{J}$ can be chosen according to an optimization problem where $p_\mathrm{mp}$ or $p_\mathrm{wc}$ can be maximized. Moreover, in line of principle the stability condition may be made less stringent by using intermediate solutions, e.g. coupling 4 sensors in two pairs and so on.
It is worth mentioning that condition (2) is the same for any kind of receiver, while condition (1) may be valid only for receivers with multi-packet reception capabilities. In this sense, there are cases where stability is guaranteed if multi-packet reception is enabled, while it may be not for a standard receiver. This improvement is clearly visible with SIC since it enhances the probability of receiving multiple packets simultaneously, and so also $p_\mathrm{mp}$: e.g., for a system with a single unstable eigenvalue $\lambda(A)$, $\mathcal{J}=\{1,2\}$, $\Ptx_{i,\mathrm{max}}=1$, $s_i=1$, $\sigma^2=0.1$, it can be numerically derived that condition (2) requires $\lambda(A)<1.6$, whereas condition (1) without SIC is $\lambda(A)<1.05$, while condition (1) with SIC is $\lambda(A)<2.6$. 
 
We are now ready to state the existence of a (stationary) optimal policy for problem (\ref{eq:problem}). Define the Value function
\begin{equation}
V_\beta(P)= \min_U J(U,P)
\end{equation}
and let $\mathcal{X}^+(P)$ denote the finite set of possible error covariance matrices that can be reached in one step from~$P$
\begin{align*}
&\mathcal{X}^+(P) = \left\{z\in \mathcal{X} : \vphantom{\sum_{i=1}^{N}} \right.\nonumber\\ 
&\quad\left.z\!=\!A\!\left(\!P^{-1} \!+\! \sum_{i=1}^{N}\gamma_iC'_iR_i^{-1}C_i\!\right)^{-1}\!A'
\!+\!Q, \ \gamma_i\in \{0,1\} \right\}
\end{align*} 
Then we have the following result.
\begin{prop}\label{th:stat}
Under the hypotheses of Lemma \ref{th:stability}, there exists an optimal infinite-horizon policy that is stationary. The optimal action $u^*(P)$ can be found by solving the optimality equation 
\begin{equation}\label{eq:bellman}
V_\beta(P)=\mathcal{C}(P,u^*(P)) + \beta \!\sum_{z\in \mathcal{X}^+(P)}\! V_\beta(z)\bm{P}(z|P,u^*(P))
\end{equation}
\end{prop}\vspace{-15pt}
\begin{pf} See Appendix A. \end{pf}\vspace{-15pt}

The previous proposition ensures that the problem admits a feasible solution and allows us to limit our search on the set of stationary policies. To this end, there are many well-known algorithms based on dynamic programming tools such as the \textit{Value Iteration algorithm} and the \textit{Relative Value Iteration algorithm} \cite{bertsekas2000dynamic}. 
These algorithms find the fixed point $V_\beta(P)$ that satisfies \eqref{eq:bellman} by iteratively finding $u^*(P)$ that minimizes the right hand side for the value of $V_\beta(P)$ at the current iteration, until convergence. 
To manage the continuous state space, it can be discretized. See Sec. \ref{sec:sim} for details on this aspect.

\section{Optimal power allocation: finite horizon}\label{sec:finpolicy}

Similarly to what is done for the infinite-horizon case, we can formulate also the finite-horizon power allocation problem. Indeed, depending on the application, the finite-horizon optimal policy may bring practical advantages, for example, when the system can be observed only for a fixed amount of time.
For an arbitrary horizon of length $K$, the problem is to find a policy $U$
\begin{equation}
U=\left\{u_1, u_2, u_3, \dots, u_K\right\}
\end{equation}
that minimizes the finite-horizon cost
\begin{align}
J\left(U, P(0)\right)= \sum_{k=0}^{K-1} \beta^k\mathcal{C}(P(k|k\!-\!1),u_k(P(k|k\!-\!1)) 
\end{align}
with $\beta\in(0,1]$. In that case, the existence of the optimal policy is straightforward since stability is not required.

We conclude the section with a surprising results for the case with one-step-ahead horizon.
\begin{prop}\label{th:optmultipkt}
Consider $K=1$ and $\mathcal{P}_i=\{0,\,\Ptx_{i,\mathrm{max}}\}$.
Assume that the sensors samples distinct outputs, namely such that $C_jC_i'\!=\!0$ for $i\neq j$. 
Then there always exists an error covariance matrix $\bar{P}$ where the optimal action $u^*(\bar{P})$ is such that $\Ptx_i=\Ptx_{i,\,\mathrm{max}}$ for any $i$.
\end{prop}\vspace*{-15pt}
\begin{pf} 	See Appendix B. \end{pf}\vspace*{-15pt}

Interestingly enough, this proposition holds in general but not for scalar systems, where all the sensors observe noisy versions of the same quantity and so the assumption $C_jC_i'=0$ does not hold. In fact, as it has been shown in our conference contribution \cite{pezzutto2020transmission}, there are configurations of the parameters for which there does not exist an error variance such that multiple simultaneous transmissions are optimal. Conversely, the previous proposition applies, among the others, for the notable case where each sensor observes a different component of the state.

\section{Special case: decoupled systems}\label{sec:decsys}

\begin{figure}
	\centering
	\begin{tikzpicture}	

	\coordinate (sys1) at (-0.1,0.5);
	\coordinate (sys2) at (0,0);
	\coordinate (sysN) at (0.3,-0.5);
	\coordinate (sens1) at (2,0.5);
	\coordinate (tx1) at (3.7,0.5);
	\coordinate (sens2) at (2,0);
	\coordinate (tx2) at (3.7,0);
	\coordinate (txN) at (3.7,-0.5);
	\coordinate (sensN) at (2,-0.5);
	\coordinate (est) at (6.7,0);
	
	\node[draw, minimum width=1.2cm, minimum height=.5cm, line width=1pt, align=center] (SYS1) at (sys1) {$\mathrm{SYS}_1$};
	\node[] (SYS2) at (sys2) {\dots};
	\node[draw, minimum width=1.2cm, minimum height=0.5cm, line width=1pt, align=center] (SYSN) at (sysN) {$\mathrm{SYS}_N$};
	\node[draw, minimum width=1.7cm, line width=1pt, align=center] (SENS1) at (sens1) {Sensor 1};
	\node[minimum width=1.7cm, line width=1pt, align=center] (SENS2) at (sens2) {\dots};
	\node[draw, minimum width=1.7cm, line width=1pt, align=center] (SENSN) at (sensN) {Sensor $N$};
	\node[draw, minimum width=0.7cm, line width=1pt, align=center] (TX1) at (tx1) {TX};
	\node[draw, minimum width=0.7cm, line width=1pt, align=center] (TXN) at (txN) {TX};
	\node[draw, minimum width=1.7cm, minimum height=1.5cm, line width=1pt, align=center] (EST) at (est) {Estimator};
	
	\draw[dotted, line width=0.5pt] ($(SYS1.north west)+(-0.075,0.075)$)  rectangle ($(SYSN.south east)+(0.075,-0.075)$);
	
	\coordinate (switch3-START) at ($(TX1)+(0.9,0)$);
	\coordinate (switch3-CLOSED) at ($(switch3-START)+(0.5,0)$);
	\coordinate (switch3-OPEN) at ($(switch3-START)+(0.4325,0.25)$);	
	\coordinate (switch4-START) at (switch3-START |- TXN);
	\coordinate (switch4-CLOSED) at ($(switch4-START)+(0.5,0)$);
	\coordinate (switch4-OPEN) at ($(switch4-START)+(0.4325,0.25)$);	
	
	\draw[-latex] (SYS1) -- (SENS1.west) node[left, pos=0.5]{};
	\draw[-latex] (SYSN) -- (SENSN.west) node[left, pos=0.5]{ };
	\draw[-latex] (SENS1) -- (TX1) node[pos=0.5, above]{$y_1$}; 
	\draw[-latex] (SENSN) -- (TXN) node[pos=0.5, above]{$y_{N}$}; 
	\draw[-] (TX1) -- (switch3-START) node[pos=0.5, above]{}; 
	\draw[-] (switch3-START) -- (switch3-OPEN); 
	\draw[-latex] (switch3-CLOSED) --  (EST.west |- switch3-CLOSED); 
	\draw[-] (TXN) -- (switch4-START) node[pos=0.5, above]{}; 
	\draw[-] (switch4-START) -- (switch4-OPEN); 
	\draw[-latex] (switch4-CLOSED) --  (EST.west |- switch4-CLOSED); 
	
	\draw[->, line width=1pt] ($(switch3-START)+(0,0.25)$) to [in=90] ($(switch3-CLOSED)-(0.25,0.1)$);
	\draw[->, line width=1pt] ($(switch4-START)+(0,0.25)$) to [in=90] ($(switch4-CLOSED)-(0.25,0.1)$);
	\node[] (l) at ($(switch3-START)+(0.25,0.4)$) {$\gamma_1$};
	\node[] (l) at ($(switch4-START)+(0.25,-0.3)$) {$\gamma_N$};
	\node[] (l) at ($(SYS)+(4.9,0)$) {\dots};
	
	\pgfmathsetmacro{\x}{5.5}
	\pgfmathsetmacro{\y}{-0.6}
	\pgfmathsetmacro{\s}{0.6}
	\pgfmathsetmacro{\l}{1.2}
	\begin{scope}[shift={(5.5,5.25)},rotate=270]
	\draw (-1.6*\s+\x,-0.7*\s*\l+\y) .. controls (-2.3*\s+\x,-1.1*\s*\l+\y)
	and (-2.7*\s+\x,0.3*\s+\y) .. (-1.7*\s+\x,0.3*\s+\y) .. controls (-1.6*\s+\x,0.7*\s+\y)
	and (-1.2*\s+\x,0.9*\s+\y) .. (-0.8*\s+\x,0.7*\s+\y) .. controls (-0.5*\s+\x,1.5*\s+\y)
	and (0.6*\s+\x,1.3*\s+\y) .. (0.7*\s+\x,0.5*\s+\y) .. controls (1.5*\s+\x,0.4*\s+\y)
	and (1.2*\s+\x,-1*\s*\l+\y) .. (0.4*\s+\x,-0.6*\s*\l+\y) .. controls (0.2*\s+\x,-1*\s*\l+\y)
	and (-0.2*\s+\x,-1*\s*\l+\y) .. (-0.5*\s+\x,-0.7*\s*\l+\y) .. controls (-0.9*\s+\x,-1*\s*\l+\y)
	and (-1.3*\s+\x,-1*\s*\l+\y) .. cycle;
	\end{scope}	

\end{tikzpicture}\vspace{-5pt}
	\caption{Setup. Dynamics of systems ($\mathrm{SYS}_i$) are decoupled.}
	\label{fig:setup:dec}
\end{figure}
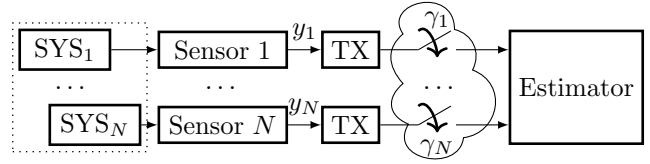

In this section, we present the analytical characterization of the optimal policy with horizon of length $K=1$ for the case of decoupled systems. Indeed, considering decoupled systems is interesting because it resembles the case where multiple independent systems are observed by a central unit. This scenario has been specifically considered e.g. by \cite{shi2012scheduling}, \cite{li2014multi}, \cite{han2017optimal}, \cite{zoppi2019transmission}.
We consider the case where each system is provided with only a (possibly multi-dimensional) sensor, as the case reported in Fig.~\ref{fig:setup:dec}, and we specialize the notation of Sec.~\ref{sec:sysmodel} as
\begin{gather}\label{eq:decsysmodel1}
A =
\left[ \begin{array}{ccc}
A_{1} & & 0 \\ & \cdots & \\ 0 & &  A_{N}
\end{array} \right] \quad
C =
\left[ \begin{array}{ccc}
C_{10} & & 0 \\ & \cdots & \\ 0 & & C_{N0}
\end{array} \right] \\ \label{eq:decsysmodel2}
Q =
\left[ \begin{array}{ccc}
Q_{1}  & & 0 \\ & \cdots & \\ 0 & &  Q_{N}
\end{array} \right] \quad
P =
\left[ \begin{array}{ccc}
P_{1} \ \, & & 0 \\ & \cdots & \\ 0 & &  P_{N} \ \,
\end{array} \right]
\end{gather}
where $A_i, Q_i, P_i \in \mathbb{R}^{n_i\times n_i}$, $C_{i0}\in\mathbb{R}^{m_i\times n_i}$, and 0 should be interpreted as a block of suitable dimension with all null entries. Denote $u=([u]_1,\dots,[u]_N)'$. 
Introduce
\begin{equation}\label{eq:opth}
\psi_i(P_i) = \mathrm{Tr} \left[A_iP_iC'_{i0}(C_{i0}P_iC'_{i0}+R_i)^{-1}C_{i0}P_iA'_i\right]
\end{equation}
and $\psi(P)=(\psi_1(P_1),\dots,\psi_N(P_N))$. Roughly speaking, it is the term by which the trace of estimation error covariance is reduced if a measurement is received. In this sense, it can be seen as the gain given by a new measurement with respect to the case where it has been lost.
\begin{prop}\label{th:decsys}
Consider a decoupled system and horizon $K=1$. Then $J(u,P)<J(v,P)$ if and only if
\begin{equation}
\sum_{i=1}^{N} \psi_i(P_i)p_i(u)  + \bar{\mu} u > \sum_{i=1}^{N}  \psi_i(P_i)p_i(v) + \bar{\mu} v  
\end{equation}
\end{prop}\vspace*{-15pt}
\begin{pf} See Appendix C. \end{pf}\vspace*{-5pt}

Intuitively, since $\psi_i(P)>0$ and $p_i$ is monotonically increasing with respect to the power allocated to the $i$-th sensor, the optimal policy allocates more power to the sensor with the largest $\psi_i(P)$. 
On the other hand, since $p_i$ is monotonically decreasing with respect to the power allocated to the $j$-th sensor, the optimal policy looks for a balance of the allocated powers weighted by $\psi_i(P)$.

As an immediate consequence, the previous proposition shows that the optimal policy has a threshold-like behaviour with respect to the transformation $\psi(P)$ of $P$. Indeed, for any possible control action $u$, we can set a system of $|\mathcal{U}|$ linear inequalities in the variables $\psi_i(P_i)$ obtaining the region where the control action $u$ is optimal. It follows that the optimal action can be expressed using an indicator function and can be implemented using a finite memory to store the thresholds.

Moreover, since $\psi_i(P_i)$ is monotonically increasing in $P_i$, we can obtain the following characterization of the optimal policy with respect to the error covariance $P_i$.
\begin{prop}\label{th:decsysmonotone}
Consider a decoupled system and horizon $K=1$. If $\bar{P}_i\geq P_i$ and $\bar{P}_j=P_j$ for $j\neq i$, then for the optimal action $u^*$ it holds that 
$[u^*(\bar{P})]_i \geq [u^*(P)]_i.$
\end{prop}\vspace*{-15pt}
\begin{pf}
See Appendix C.
\end{pf}\vspace*{-15pt}

The following corollary clarifies the easiest case with only two sensors and two power levels. The proof is omitted since it consists of standard computations.

\begin{cor}\label{th:decsys2lev}
Consider a decoupled system, $K=1$, $N=2$, and $\mathcal{P}_i=\{0,\,\Ptx_{i,\mathrm{max}}\}$. Define
\begin{align*}
&\!S_{00}\!=\!\left\{ P:\psi_1(P_1)\!<\!\frac{\mu}{p_{10}(1,\!0)},\, \psi_2(P_2)\!<\!\frac{\mu}{p_{01}(0,\!1)} \right\}\\
&S_{11}\!=\!\!\!
\begin{multlined}[t]
\left\{ P \!:\! (p_{10}(1,\!0)\!-\!p_{1}(1,\!1))\psi_1(P_1) \!+\!\mu \!<\! p_{2}(1,\!1)\psi_2(P_2) \right.\\ \left. \text{and } (p_{01}(0,\!1)\!-\!p_{2}(1,\!1))\psi_2(P_2) \!+\! \mu \!<\! p_{1}(1,\!1)\psi_1(P_1)\right\}
\end{multlined}\\
&S_{10}\!=\!\left\{ P:p_{10}(1,\!0)\psi_1(P_1)\!>\! p_{01}(0,\!1)\psi_2(P_2)\right\} \!\setminus\! \left\{S_{00} \!\cup\! S_{11}\right\}\\
&S_{10}\!=\!\left\{ P:p_{10}(1,\!0)\psi_1(P_1)\!<\! p_{01}(0,\!1)\psi_2(P_2)\right\} \!\setminus\! \left\{S_{00} \!\cup\! S_{11}\right\}
\end{align*}
Then the optimal action $u^*(P)$ is 
\begin{equation}
u^*(P)=
\begin{cases}
(0,\,0) &\text{ if } P \in S_{00}\\
(0,\,\Ptx_{2,\mathrm{max}}) &\text{ if }  P \in S_{01}\\
(\Ptx_{1,\,\mathrm{max}},0) &\text{ if }  P \in S_{10}\\
(\Ptx_{1,\mathrm{max}},\,\Ptx_{2,\mathrm{max}}) &\text{ if }  P \in S_{11}
\end{cases}
\end{equation}
\end{cor}

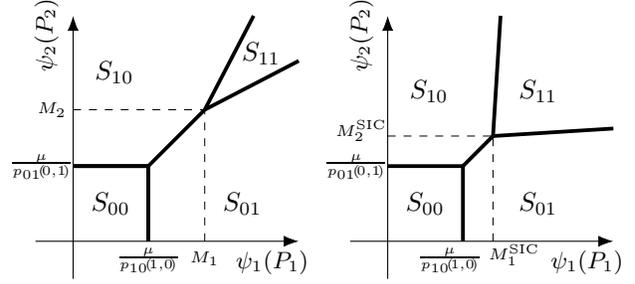
\begin{figure}[t]
	\hspace*{-5pt}
	\begin{minipage}{0.4\linewidth}
		\centering
		\begin{tikzpicture}
	\coordinate (x0) at (-0.5,0);
	\coordinate (y0) at (0,-0.5);
	\coordinate (x) at (3,0);
	\coordinate (y) at (0,3);
	
	\coordinate (r1start) at (1,0);
	\coordinate (r1end) at (1,1);
	\coordinate (r2start) at (0,1);
	\coordinate (r3end) at (1.75,1.75);
	\coordinate (r3bott) at (1.75,0);
	\coordinate (r3left) at (0,1.75);
	\coordinate (r4end) at (2.4,3);
	\coordinate (r5end) at (3,2.4);
	
	\draw[-{Latex[length=2mm, width=1.5mm]}, thin] (x0) --  (x) node[pos=0.9, below] {\footnotesize $\psi_1(P_1)$};
	\draw[-{Latex[length=2mm, width=1.5mm]}, thin] (y0) -- (y) node[pos=0.9, left, rotate=90, yshift=10pt, xshift=20pt] {\footnotesize $\psi_2(P_2)$};
	
	\draw[-, line width=0.5mm] (r1start) -- (r1end);
	\draw[-, line width=0.5mm] (r2start) -- (r1end);
	\draw[-, line width=0.5mm] (r1end) -- (r3end);
	\draw[-, line width=0.5mm] (r3end) -- (r4end);
	\draw[-, line width=0.5mm] (r3end) -- (r5end);
	\draw[-, line width=0.5mm] (r1start) -- (r1end) node[pos=0, yshift=-7pt] {\tiny $\frac{\mu}{p_{10}\!(\!1,0\!)}$};
	\draw[-, line width=0.5mm] (r2start) -- (r1end) node[pos=0, xshift=-12pt] {\tiny $\frac{\mu}{p_{01}\!(\!0,1\!)}$};
	
	\draw[dashed, line width=0.1mm] (r3bott)-- (r3end) node[pos=0, yshift=-6.5pt] {\tiny $M_1$} ;
	\draw[dashed, line width=0.1mm]  (r3left)  -- (r3end) node[pos=0, xshift=-8pt] {\tiny $M_2$} ;

	\node[](l) at (0.5,0.5){$S_{00}$};
	\node[](l) at (0.55,2.25){$S_{10}$};
	\node[](l) at (2.25,0.5){$S_{01}$};
	\node[](l) at (2.5,2.5){$S_{11}$};
	
\end{tikzpicture}
	\end{minipage}\hspace*{20pt}
	\begin{minipage}{0.45\linewidth}
		\centering
		\begin{tikzpicture}
	\coordinate (x0) at (-0.5,0);
	\coordinate (y0) at (0,-0.5);
	\coordinate (x) at (3,0);
	\coordinate (y) at (0,3);
	
	\coordinate (r1start) at (1,0);
	\coordinate (r1end) at (1,1);
	\coordinate (r2start) at (0,1);
	\coordinate (r3end) at (1.4,1.4);
	\coordinate (r3bott) at (1.4,0);
	\coordinate (r3left) at (0,1.4);
	\coordinate (r4end) at (1.5,3);
	\coordinate (r5end) at (3,1.5);
	
	\draw[-{Latex[length=2mm, width=1.5mm]}, thin] (x0) --  (x) node[pos=0.9, below] {\footnotesize $\psi_1(P_1)$};
	\draw[-{Latex[length=2mm, width=1.5mm]}, thin] (y0) -- (y) node[pos=0.9, left, rotate=90, yshift=10pt, xshift=20pt] {\footnotesize $\psi_2(P_2)$};
	
	\draw[-, line width=0.5mm] (r1start) -- (r1end) node[pos=0, yshift=-7pt, xshift=-5pt] {\tiny $\frac{\mu}{p_{10}\!(\!1,0\!)}$};
	\draw[-, line width=0.5mm] (r2start) -- (r1end) node[pos=0, xshift=-12pt] {\tiny $\frac{\mu}{p_{01}\!(\!0,1\!)}$};
	\draw[-, line width=0.5mm] (r1end) -- (r3end);
	\draw[-, line width=0.5mm] (r3end) -- (r4end);
	\draw[-, line width=0.5mm] (r3end) -- (r5end);

	\draw[dashed, line width=0.1mm] (r3bott)-- (r3end) node[pos=0, yshift=-5pt, xshift=8pt] {\tiny $M^\textrm{SIC}_1$} ;
	\draw[dashed, line width=0.1mm]  (r3left)  -- (r3end) node[pos=0, xshift=-10pt, yshift=2pt] {\tiny $M^\textrm{SIC}_2$} ;

	\node[](l) at (0.5,0.5){$S_{00}$};
	\node[](l) at (0.55,2){$S_{10}$};
	\node[](l) at (2,0.5){$S_{01}$};
	\node[](l) at (2,2){$S_{11}$};
	
\end{tikzpicture}
	\end{minipage}\vspace{-5pt}
	\caption{Optimal policy without SIC (left) and with SIC (right)
		with $M_1\!=\!\mu p_{01}(0,\!1)/p_e$ and $M_2\!=\!\mu p_{10}(1,\!0)/p_e,$ where $p_e=p_1(1,\!1)p_{10}(1,\!0) + p_2(1,\!1)p_{01}(0,\!1) - p_{10}(1,\!0)p_{01}(0,\!1)$ and similarly with SIC.
		It is easy to show that $M_1\!>\!M^\mathrm{SIC}_1$, $M_2\!>\!M^\mathrm{SIC}_2$, and $S_{11}$ is larger with SIC.}
	\label{fig:optregion}
\end{figure}

The previous corollary identifies 4 simple regions given by the optimality conditions provided by Proposition \ref{th:decsys}. Figure \ref{fig:optregion} reports the typical shape of the regions for the case with and without SIC. 
The reader can find an analogy between Fig.~\ref{fig:arrivalregion}, that depicts the received powers and the corresponding value of the arrival variable, and Fig.~\ref{fig:optregion}, that depicts the values of the functions $\psi_i(\cdot)$ of the error covariance and the corresponding optimal action. 
Remarkably, $R_{00}$ and $S_{00}$ are the same with and without SIC, while $R_{11}$ and $S_{11}$ without SIC are strictly included in $R_{11}$ and $S_{11}$ with SIC.
Once again, we can see the improvement given by SIC: as the arrival probability was increased in Fig.~\ref{fig:arrivalregion}, the mean error covariance is improved in Fig.~\ref{fig:optregion}, at the price of higher mean power consumption.
Unfortunately, in general, it is not possible to analytically associate the error covariance matrix $P$ to the optimal action $u^*(P)$ without passing through $\psi_i(\cdot)$. However, this is possible if $A_i$ is scalar, where it is easy to invert the functions $\psi_i(\cdot)$. In that case, the previous proposition specializes to Proposition 5 in \cite{pezzutto2020transmission}.

\section{Simulations}\label{sec:sim}

\textit{Algorithm Implementation.}
To obtain the optimal infinite-horizon policy, we rely on the Value iteration algorithm \cite{bertsekas2000dynamic}. Even though it can be applied to MDPs with general state and action spaces, they are commonly discretized for numerical purposes.
In our case, the action space $\mathcal{U}$ is already finite, while the state space $\mathcal{X}$ is the infinite-dimensional set of positive semidefinite matrices, for which a universal quantization mechanism does not exist.
In this paper, we propose to obtain a discrete set as follows: first we collect 
multiple sample paths generated with random arrival processes and random initial conditions, then we cluster them into $D$ different clusters based on the Frobenius norm mimicking what is done by the k-means algorithm for vector quantization, finally we obtain the set $\mathcal{X}_d$ by including the cluster centroids. In this way, $D$ can be tuned to obtain a trade-off between the accuracy of the quantization and the computational burden, while the focus is on the matrices given by the actual updates of the covariance matrix instead of exploring the whole space $\mathcal{X}$.
More sophisticated methods are out of the scope of this paper.
Based on the discretized state space $\mathcal{X}_d$, we find the optimal policy off-line as reported in Algorithm \ref{alg}. It consists of two steps: first, the finite set $\mathcal{X}^+_d(P)$ of the discretized error covariance matrices that can be reached in one step from $P$ is computed for any $P\in\mathcal{X}_d$, then, an iterative procedure is carried out to find the fixed point of equation \eqref{eq:bellman} based on the Value Iteration algorithm. The optimal policy consists of a map from any matrix in $\mathcal{X}_d$ to the corresponding optimal action and it can be stored in a lookup table. At time instant $k$, we find the matrix belonging to $\mathcal{X}_d$ that minimizes the distance to $P(k+1|k)$ and we use the corresponding optimal action.

\begin{algorithm}
\caption{Get optimal policy}\label{alg}
\begin{algorithmic}
	\State $V_0 \gets \textrm{initialize randomly}$
	\For{$P \in \mathcal{X}_d$}
	\State $\mathcal{X}_d^+(P)=\emptyset$
	\For{$\nu=(\nu_1,\dots,\nu_N)\in \{0,1\}^N$}
	\State $P^+=A(P^{-1} + \sum_{i} \nu_i C_i'R_i^{-1}C_i)^{-1}A' + Q$
	\State $P^+_d=\arg\min_{z\in \mathcal{X}_d} |\!| z - P^+|\!| $
	\State $\mathcal{X}_d^+(P)=\mathcal{X}_d^+(P) \cup \{P^+_d\}$
	\For{$u\in\mathcal{U}$}
		\State $\mathbf{P}(P_d^+|P,u)=\mathbf{P}(\gamma=\nu|\Ptx=u)$
		\State \hspace{2pt} where $\gamma\!=\!(\gamma_1, ...\, ,\gamma_N)$ has distribution \eqref{eq:def:prob}
	\EndFor
	\EndFor
	\EndFor
	\For{$k\geq 0$}
	\For{$P \in \mathcal{X}_d$}
		\State $V_{k+1}(P)\!=\!\min\limits_{u\in\mathcal{U}} \mathcal{C}(P,u) +\!\!\! \sum\limits_{z\in\mathcal{X}_d^+(P)} \!\!\! \beta V_{k}(z)\mathbf{P}(z|P,u)$
		\State \begin{equation*}
		\begin{multlined}[t]
			U_{k+1}(P)=\arg\min\limits_{u\in\mathcal{U}}
			\mathcal{C}(P,u) \hspace{45pt} \\ + \!\!\textstyle\sum\limits_{z\in\mathcal{X}_d^+(P)} \!\!\beta V_{k}(z)\mathbf{P}(z|P,u) 
            \end{multlined}
		\end{equation*}
	\EndFor
	\If{convergence reached}
		\Return $U^*=U_{k+1}$
	\EndIf
	\EndFor
\end{algorithmic}
\end{algorithm}
\vspace*{5pt}

\textit{Assessment of the proposed strategies.} 
Since multi-packet reception is naturally applicable in control of multi-agent systems, we consider a system featuring 2 identical drones that communicate their own positions to a remote control unit.
The use of multiple drones is interesting for many different applications like environmental monitoring and object transportation.
Accordingly to the notation of \eqref{eq:decsysmodel1}-\eqref{eq:decsysmodel2}, the required parameters are chosen as
\begin{gather*}
A_1 =A_2=
\left[ \begin{array}{cc}
1 & T \\ 0 &  1
\end{array}\right] \quad
C_{10} = C_{20} =
\left[ \begin{array}{cc}
1&   0
\end{array}\right] \quad
Q = 0.1I
\end{gather*}
where $T=0.1\,\mathrm{s}$ is the sampling period. This model is used also for other systems like ground robots or simple vehicles.
It is reasonable to assume that both on-board transmitting systems are identical, and, if the drones are close to each other with respect to the central unit without obstacles on the line-of-sight, also the channels can be considered identical. For this reason we set 
\begin{gather*}
\Ptx_{1,\mathrm{max}} = \Ptx_{2,\mathrm{max}}  = 1 \quad  s_1 =  s_2 = 1 \quad \sigma^2=0.1
\end{gather*}
where we have normalized the powers and the channel power gains. We set $M=4$, $\alpha=0.75$ and $\beta=0.9$.

\begin{figure}[t]
	\begin{minipage}{\linewidth}
		\centering
		\includegraphics[width=\linewidth]{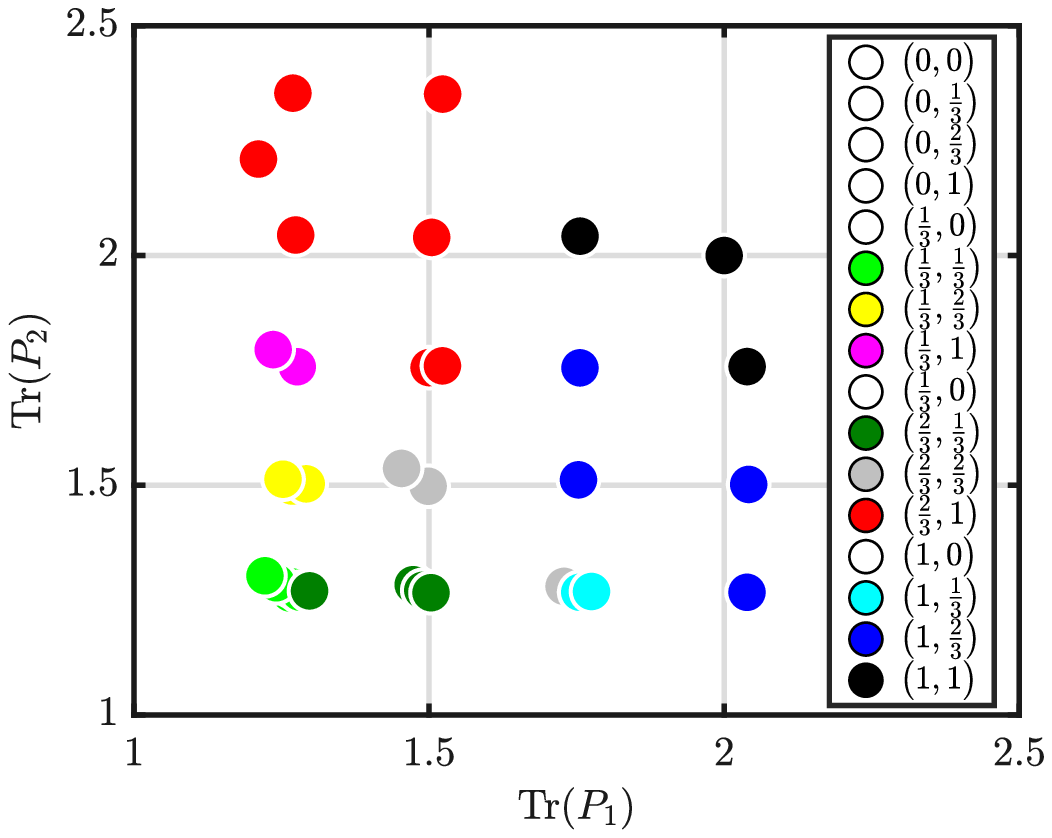}
	\end{minipage}\vspace*{-15pt}
	\begin{minipage}{\linewidth}
		\centering
		\includegraphics[width=\linewidth]{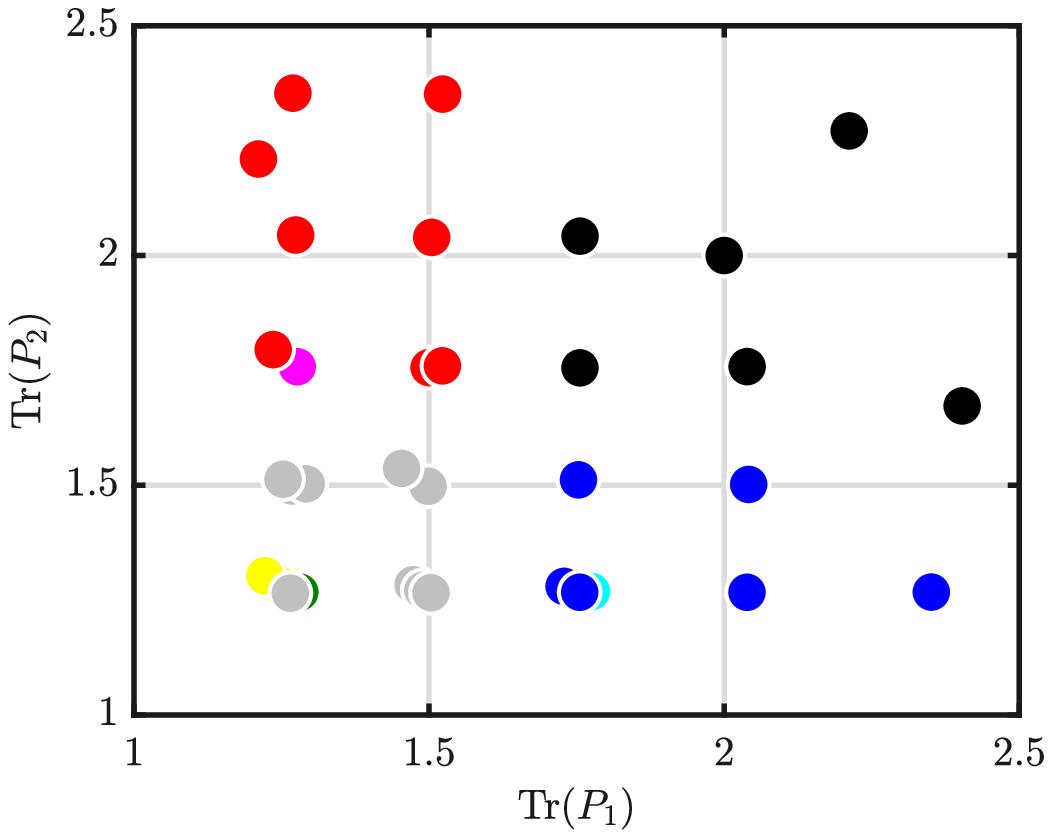}
	\end{minipage}\vspace*{-10pt}
	\caption{Optimal policy with SIC ($\mu=0.1$). 
		Top panel: $K=1$. Bottom panel: $K=\infty$. System: two drones. In the legend, no colors are assigned to actions in $\mathcal{X}_d$ not used.}
	\label{fig:policy}
\end{figure}

We start by reporting a pictorial representation of the optimal policy with SIC.
To this end, we may use the function $\psi(P)$ defined in \eqref{eq:opth}. Indeed, using the function $\psi(P)$, each covariance matrix $P$ can be mapped to a point of the plane $\psi_1(P_1) \times \psi_2(P_2)$, where $P_i$ is the error covariance matrix relative to subsystem $i$. Then, using the relation in Proposition \ref{th:decsys}, the plane $\psi_1(P_1) \times \psi_2(P_2)$ can be divided into a finite number of subregions, each of them associated to a specific optimal action, producing a plot similar to Fig.~\ref{fig:optregion}. 
Here, however, we prefer to represent the plane $\mathrm{Tr}(P_1)\times \mathrm{Tr}(P_2)$, where $\mathrm{Tr}(P_i)$ is the trace of $P_i$, instead of the plane $\psi_1(P_1) \times \psi_2(P_2)$. In fact, $\mathrm{Tr}(P_i)$ gives an immediate understanding of the quality of the current estimate of the state of subsystem $i$: the higher $\mathrm{Tr}(P_i)$, the worse the knowledge on subsystem $i$ is.
The optimal one-step-ahead policy and the optimal infinite-horizon policy with SIC are reported in the top and bottom panels of Fig.~\ref{fig:policy}, respectively. In the two plots, the points are not uniformly distributed because they are taken from the discretized state space $\mathcal{X}_d$, which is not distributed over the whole space. 
In the top panel, we can see that, for a fixed $\mathrm{Tr}(P_2)$, the optimal one-step-ahead power allocated to sensor 1 is not decreasing with respect to $\mathrm{Tr}(P_1)$. The same holds for sensor 2. Similarly, the total power increases by moving along a straight line starting from the origin. Since larger $P_i$ entails larger $\mathrm{Tr}(P_i)$, this is expected from Proposition \ref{th:decsysmonotone}, which states that the optimal one-step-ahead transmission power of sensor $i$ increases for larger $P_i$ for a fixed $P_j$.
As it can be seen in the bottom panel, the optimal infinite-horizon policy has similar features. Differences are present especially in the bottom left corner of the plane: the optimal action for error covariances $P$ such that $\mathrm{Tr}(P_1)\simeq1.25$, $\mathrm{Tr}(P_2)\simeq1.5$ shifts from $\left(\frac{1}{3},\frac{2}{3}\right)$ to $U^*=\left(\frac{2}{3},\frac{2}{3}\right)$, and the same happens for $\mathrm{Tr}(P_1)\simeq1.5$, $\mathrm{Tr}(P_2)\simeq1.25$, while the set of error covariances with $\mathrm{Tr}(P_1)\simeq\mathrm{Tr}(P_2)\simeq1.25$ is split into three different regions, and the most conservative action $\left(\frac{1}{3},\frac{1}{3}\right)$ is no more optimal for any considered covariance. It is interesting to stress that the optimal policy allocates non-null power to both sensors for any $P\in\mathcal{X}_d$, namely multi-packet reception is always used.

\begin{figure}[t]
	\begin{minipage}{\linewidth}
		\centering
		\includegraphics[width=\linewidth]{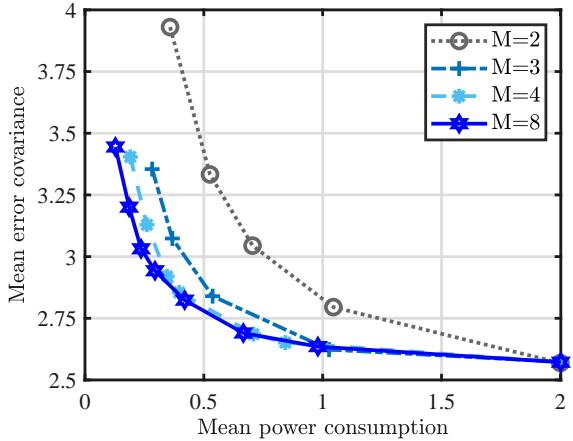}
	\end{minipage}\vspace*{-15pt}
	\begin{minipage}{\linewidth}
		\centering
		\includegraphics[width=\linewidth]{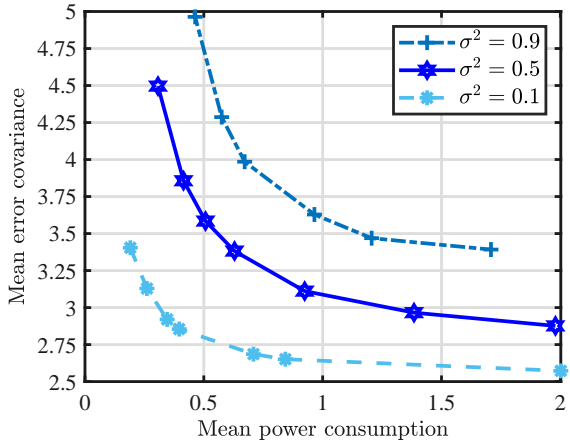}
	\end{minipage}\vspace*{-10pt}
	\caption{Assessment of the optimal infinite-horizon policy with SIC for different numbers of available power levels $M$ (top panel) and  noise $\sigma^2$ (bottom panel). System: two drones.}
	\label{fig:noise}
\end{figure}

\begin{figure}[t]
	\centering
	\includegraphics[width=\linewidth]{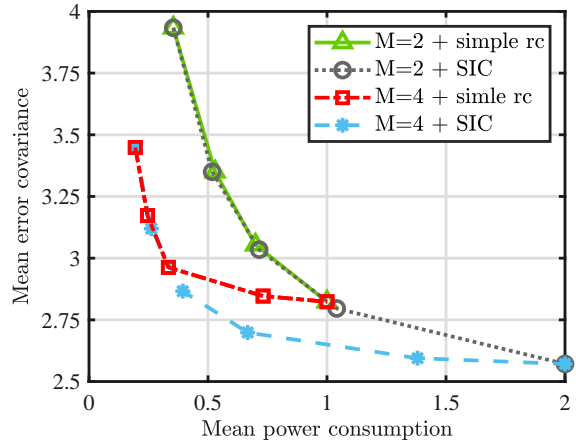}\vspace*{-10pt}
	\caption{Assessment of the optimal infinite-horizon policies for different receivers (with and without SIC) and different numbers of available power levels $M$. System: two drones.}
	\label{fig:cov-vs-tx}
\end{figure}

We now focus on the performances of the proposed policies. 
The following plots report the mean trace of the error covariance matrix for different mean power consumptions (normalized over the consumption for a single transmission at the maximum power).
Such a plot allows comparing different policies (or the same policy with different parameters) in a fair way by providing the different estimation performances achieved with the same power consumption.  
To obtain the following plots, we set different values of $\mu$ and we simulate the system evolution over a long time horizon under the corresponding policy.
To assess how the optimal infinite-horizon policy with SIC behaves for different hardware and network conditions, we vary the number of available transmission power levels (Fig.~\ref{fig:noise}, top panel) and the background noise (Fig.~\ref{fig:noise}, bottom panel).
In the top panel of Fig.~\ref{fig:noise} we can see that increasing the number of levels $M$ improves the estimation quality for the same mean power consumption. On one hand, the improvement is particularly clear for low power consumptions, where we can see that, with more transmission levels, the energy is more efficiently employed. On the other hand, the improvement tends to saturate when increasing the number of levels, especially when the energy bound is not stringent (right part of the plot): mean error covariance is identical for a mean power consumption of 2 with any considered value of $M$, and for a mean power consumption higher than 0.4 for $M=4$ and $M=8$. 
In the bottom panel of Fig.~\ref{fig:noise}, we can see that with better network conditions, i.e. lower background noise $\sigma^2$, the estimation quality is improved for the same mean power consumption.
In general, a lower error covariance can be achieved with lower $\sigma^2$ but some error covariances can be achieved also with higher $\sigma^2$ at the price of higher mean power consumptions. For instance, a mean error covariance of 3.4 can be achieved with a mean power consumption of 0.2 with $\sigma^2=0.1$, of 0.6 with $\sigma^2=0.5$, and of 1.7 with $\sigma^2=0.9$.

We now compare the performances of the infinite-horizon optimal policy for both the receivers, namely with and without SIC, and for transmitters with different numbers of available power levels, specifically $M=2$ and $M=4$. Results are reported in Fig. \ref{fig:cov-vs-tx}. 
We can see that, allowing more than 2 power levels, the mean error covariance is smaller. The difference is particularly evident when we impose a stringent constraint on the power consumption (left part of the plot). 
The performance without SIC is similar to the counterpart with SIC when the mean power consumption is small. This is due to the fact that communications are highly penalized so that simultaneous transmissions are selected for error covariances that the system reaches less often.
The difference becomes more evident when the energy constraint is less stringent: with $M=4$, we achieve a reduction of the mean trace of error covariance up to  10\%.
Note that in the case of SIC, both with $M=2$ and $M=4$, the optimal policy tends to be more aggressive in terms of power allocation: indeed, for small $\mu$, with SIC, simultaneous transmissions at the maximum power are often scheduled, while in the case without SIC, even with $\mu=0$ the mean power consumption is equal to $1$. This is due to the fact that alternating transmissions from sensor 1 to sensor 2 are preferred because the loss probability with simultaneous transmissions is high.
Presented results indicate the importance of SIC in order to always optimally allocate the available power when it is limited (left part of the plot), and to further decrease the mean error covariance when requirements on power consumption are not stringent (right part).

\begin{figure}[t]
	\centering
	\includegraphics[width=\linewidth]{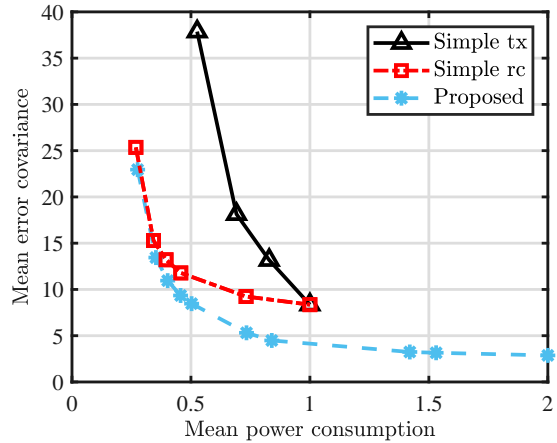}\vspace*{-10pt}
	\caption{Comparison between existing algorithms ($K=1$). Simple tx \cite{leong2016sensor}. Simple rc \cite{li2019power}. Proposed: $M=4$ with SIC. System: two pendulum-on-a-cart system.}
	\label{fig:cov-vs-tx-pend}
\end{figure}

\textit{Comparison with other strategies.} To validate the proposed algorithm we compare it with other existing algorithms that rely on simpler hardware. In particular, we consider the simplest solution studied in \cite{leong2016sensor} where transmitters have only two power levels and the receiver has no multi-packet reception capabilities (only one sensor is scheduled at a given time). We will refer to it as ``Simple tx''. 
Then we consider a more advanced solution adapted from \cite{li2019power} where the transmitters have 4 power levels and the receiver is able to decode multiple simultaneous packets but does not implement SIC. We will refer to it as ``Simple rc''. 
Finally, we consider the proposed policy as derived in Sec. \ref{sec:finpolicy} with SIC and $M=4$. 
We carry out the test on a more extreme case where the system consists of two pendulum-on-a-cart. We consider the discretization with $T=0.01\, \mathrm{s}$ of the linearized continuous-time system 
\begin{equation*}
A_1=A_2=\left[ \begin{array}{cc}
0 & 1 \\ g/\ell & 0
\end{array}\right]
\quad
C_{10}=C_{20}=\left[ \begin{array}{cccc}
1 & 0 
\end{array}\right]
\quad Q=0.1I
\end{equation*}
where g is the gravitational acceleration and $\ell=0.2$, equivalent to the case of a rigid rod with length $15\,\mathrm{cm}$. This model can capture the simplified dynamics of wheeled robots transporting an object. Differently from the system with two drones, the dynamics are unstable, so the estimation and the power allocation problem are more challenging. Communication parameters are set as for the previous example.
Results are reported in Fig. \ref{fig:cov-vs-tx-pend}.
We can see that the proposed strategy outperforms the existing algorithms. In particular, the difference with respect to Simple tx is evident when the energy requirements are strict, i.e., when the mean power consumption is small. On the other hand, the difference with respect to Simple rc is the largest for a mean power consumption equal to 1, for which the mean error covariance achieved by the proposed strategy using SIC is  half of the existing algorithm. 
Note that Simple rc achieves the same error covariance of Simple tx and it is not able to go beyond a mean power consumption of 1, namely a single transmission at the maximum power at each time step. 
Conversely, with SIC, it is convenient to have simultaneous transmissions to get information from both the systems more often. This shows that simultaneous transmissions are really advantageous with the concurrent implementation of SIC.

\section{Conclusions}

In this paper we explore the power allocation problem for remote estimation in the case where multiple incoming packets from simultaneous transmissions can be decoded by the receiver. We show the existence of a stationary optimal policy for the infinite-horizon problem and we show that the optimal one-step-ahead policy has a threshold behavior with respect to a scalar transformation of the error covariance if the system is decoupled. Numerical examples show the improvements of the proposed strategy, especially when SIC is implemented.
The proposed framework is general and allows for several extensions. 

\appendix

\section{Proofs of Sec. \ref{sec:infpolicy}}
\begin{pf*}{\textbf{PROOF of Lemma \ref{th:stability}}}
We first assume that condition (1) holds. Consider a policy $\bar{U}$ where at each time instant all the sensors belonging to the set $\mathcal{J}$ transmit.
We can define the operator 
\begin{equation*}
\bar{g}(X)=AXA' + Q - p_\mathrm{mp}AXC'_\mathcal{J}(C_\mathcal{J}XC'_\mathcal{J}+R)^{-1}C_\mathcal{J}XA'
\end{equation*}
and the sequence
\begin{equation*}
\bar{P}_{k+1}=\bar{g}(\bar{P}_k) \text{ from } \bar{P}_1=P(1|0)
\end{equation*}
We show that $\mathbb{E}[P(k|k-1)]\leq \bar{P}_{k}$ under the proposed policy by inductive argument. To this end, assume that $\mathbb{E}[P(k|k-1)]\leq \bar{P}_{k}$ for an arbitrary $k>0$.
Then we have
\begin{align*}
\mathbb{E}&[P(k+1|k)]\\
&=\mathbb{E}\left[A\left(P(k|k-1)^{-1}+\sum\nolimits_{i}\gamma_iC'_iR_i^{-1}C_i\right)^{-1}A'+Q\right]\\
&\leq \begin{multlined}[t]
\mathbb{E}\!\left[ \!A\!\left(P(k|k\!-\!1)^{-1} \!\!+\!\! \sum\nolimits_{i\in\mathcal{J}}\!C'_iR_i^{-1}C_i\right)^{-1}\!\! A'\!+\! Q \right]\!p_\mathrm{mp}  \\ +\mathbb{E}[AP(k|k\!-\!1)A'\!+\!Q](1\!-\!p_\mathrm{mp})\end{multlined}\\
&\leq A\bar{P}_{k}A' + Q  -p_\mathrm{mp}A\bar{P}_{k}C_\mathcal{J}'(C_\mathcal{J}\bar{P}_{k}C'_\mathcal{J}\!+\!R_\mathcal{J})^{-1}C_\mathcal{J}\bar{P}_{k}A' \\
&= \bar{P}_{k+1}
\end{align*}
where we used the Matrix Inversion Lemma and the Jensen inequality. We can conclude that $\mathbb{E}[P(k|k-1)]\leq \bar{P}_{k}$ for any $k\geq 0$. In addition from \cite{sinopoli2004kalman} we know that if $p_\mathrm{mp}>\Lambda(A)$ there exists a $M_{P(0)}$ such that $\bar{P}_{k}\leq M_{P(0)}$. It follows that $\mathbb{E}[P(k|k-1)]\leq M_{P(0)}$ proving the claim. \\

We now assume that condition (2) holds. Consider a periodic policy $\bar{U}$ where at each time instant a sensor $i\in\mathcal{J}$ transmits at the maximum power, while other sensors do not communicate. If we relabel the sensors so that $\mathcal{J}=\{1,2,\dots,L\}$, a possible policy is
\begin{equation*}
[\bar{u}_k(P(k|k-1))]_i =
\begin{cases}
\Ptx_{i,\,\mathrm{max}} &\text{if } i\!\in\!\mathcal{J} \text{ and } \exists \ell\!\in\!\mathbb{N}:k\!=\!L\ell\!+\!i\\
0 &\text{otherwise}
\end{cases}
\end{equation*}
where $[u]_i$ is the $i$-th entry of the vector $u$. For sake of clarity, we prove the lemma for $\mathcal{J}=\{1,2,3,4\}$ but the general case can be proved analogously. Introduce the following operators
\begin{align*}
h(X) \,
&= AXA' + Q \\
g_i(X) 
&= X - p_iXC_i'(X+C_iR_iC_i')^{-1}C_iX \\
&= (X^{-1} + p_iC_iR_i^{-1}C_i)^{-1}\\
\tilde{g}_i(X) 
&= X - p_{\mathrm{wc}}XC_i'(X+C_iR_iC_i')^{-1}C_iX \\
&= (X^{-1} + p_{\mathrm{wc}}C_iR_i^{-1}C_i)^{-1}.
\end{align*}
Under the proposed periodic policy we have
\begin{align*}
\mathbb{E}[P(&4k+4|4k-3)] \\
&\leq h\circ g_4 \circ h\circ g_3 \circ h\circ g_2 \circ h\circ g_1 \left(\mathbb{E}[P(4k|4k-1)] \right) \\
&\leq h\circ \tilde{g}_{4} \circ h\circ \tilde{g}_{3} \circ h\circ \tilde{g}_{2} \circ h\circ \tilde{g}_{1} \left(\mathbb{E}[P(4k|4k-1)] \right) \\
&\leq h\circ  h \circ h\circ h \circ  \tilde{g}_{4} \circ \tilde{g}_{3} \circ \tilde{g}_{2} \circ \tilde{g}_{1} \left(\mathbb{E}[P(4k|4k-1)] \right)
\end{align*}
where the first inequality holds because $h\circ g_i(X)$ is concave  \cite{sinopoli2004kalman}, the second holds because $h\circ g_i(X)$ is monotonically decreasing w.r.t. $p_i$, and the third holds if $Q>\sum_{i\in\mathcal{J}}(C'_iR^{-1}_iC_i)$ \cite{yang2011deterministic}, otherwise a similar relation can be obtained with some manipulations. 

Consider now the system $\bar{A}=A^4$, $\bar{Q}=\sum_{t=0}^{3} A^tQ$, $\bar{C}=C_\mathcal{J}$, $\bar{R}=R_\mathcal{J}$. 
We can define the operator
\begin{equation*}
\bar{g}(X)= AXA' + Q -  p_\mathrm{wc}AXC'_\mathcal{J}(C_\mathcal{J}XC'_\mathcal{J}+R)^{-1}C_\mathcal{J}XA'
\end{equation*}
and the sequence
\begin{equation*}
\bar{P}_{k+1}=\bar{g}(\bar{P}_{k}) \text{ from } \bar{P}_1=P(1|0)
\end{equation*}
We have that $\bar{g}(X)=h\circ  h \circ h\circ h \circ \tilde{g}_{4} \circ \tilde{g}_{3} \circ \tilde{g}_{2} \circ \tilde{g}_{1}(X)$ by \cite{pezzutto2020adaptive}. Similarly to what is done above, we can prove by induction that $\mathbb{E}[P(4k|4k-1)]\leq \bar{P}_{k}$ for any $k\geq 0$. It is known from \cite{sinopoli2004kalman} that if $p_\mathrm{wc}>\Lambda(\bar{A})$, then $\exists M_{P(0)}>0$ such that $\bar{P}_{k}\leq M_{P(0)}$. It follows that $\mathbb{E}[P(k|k-1)]\leq M_{P(0)}$ concluding the proof. 
\end{pf*}

\begin{pf*}{\textbf{PROOF of Proposition \ref{th:stat}}}
By Lemma \ref{th:stability} there exist a policy $\bar{U}=(\bar{u}_1,\bar{u}_2\, \dots)$ and a matrix $M_{P(0)}$ such that $\mathbb{E}[P(k|k-1)] \leq M_{P(0)}$.
It follows that 
\begin{align*}
\mathbb{E}[\mathcal{C}(P(k+1|k),\bar{u}_{k+1}(P(k+1&|k)))|P(1|0)=P]\\
&\leq \mathrm{Tr}(M_{P}) + \mu N= \overline{M}_{P}
\end{align*}
Then
\begin{align*}
	J(&\bar{U},P) \\
	&=\sum_{k=0}^{\infty}  \beta^k\mathbb{E}\left[\mathcal{C}(P(k\!+\!1|k),\bar{u}_{k+1}(P(k\!+\!1|k))) \Big| P(0) \!=\!P  \right]\\
	&\leq\sum_{k=0}^{\infty}  \beta^k \overline{M}_{P} = \frac{\overline{M}_P}{1-\beta}
\end{align*}
Finally,
\begin{equation*}
	V_\beta(P) = \inf_UJ(U,P) \leq J(\bar{U},P) \leq  \overline{M}_P/(1-\beta)  
\end{equation*}
namely the Value function is finite. Then, the existence of an optimal stationary policy follows from Theorem 1 in \cite{ritt1992optimal}. The optimality equation can be  obtained by noting that only a finite number of error covariance values of $P(k+1|k)$ can be reached from $P(k|k-1)$ in one step.
\end{pf*}

\section{Proofs of Sec. \ref{sec:finpolicy}}
For sake of simplicity, we consider the easiest case with two sensors.
Preliminarily let us denote $q_{1}=p_{10}(1,0)$ and $q_{2}=p_{01}(0,1)$, while for simplicity we denote $p_{11}=p_{11}(1,1)$, $p_{10}=p_{10}(1,1)$, $p_{01}=p_{01}(1,1)$, and $p_{00}=p_{00}(1,1)$. Then, with the new notation, according to formulas of Sec. \ref{sec:chanchar} we have
\begin{equation*}
p_{11} + p_{10} = q_{1}\frac{\lambda_2}{\lambda_2 + \alpha \lambda_1} \qquad
p_{11} + p_{01} = q_{2}\frac{\lambda_1}{\lambda_1 + \alpha \lambda_2}
\end{equation*}
Then for $\alpha\in(0,1)$ we have
\begin{multline*}
q_{1}q_{2} - q_{1}(p_{11}\!+\!p_{01}) - q_{2}(p_{11}\!+\!p_{01})=\\ q_{1}q_{2}\left(1 - \frac{\lambda_1}{\lambda_1 + \alpha \lambda_2} - \frac{\lambda_2}{\lambda_2 + \alpha \lambda_1}\right) <0
\end{multline*}
where the inequality holds because the term in parentheses in monotonically increasing with respect to $\alpha$ and equal to 0 at $\alpha=1$. We are now ready to prove the proposition.
\begin{pf*}{\textbf{PROOF of Proposition \ref{th:optmultipkt}}}
By hypothesis $C_1C_2'=0$ and without loss of generality we assume $C_1C_1'=C_2C_2'=1$. Then we can choose a set of $n-2$ vectors $v_3,v_4,\dots,v_n$ (row for convenience) such that $C_iv_j'=0$, $i=1,2$, $j=3,4,...,n$,  $v_\ell v_j'=0$, $\ell,j=3,4,...,n$, and $v_j v_j'=1$, $j=3,4,...,n$, obtaining an orthonormal basis of the space $\mathbb{R}^n$. Using such a basis we define the set $\mathcal{X}_\sigma$ of symmetric positive definite matrices of the form
\begin{equation*}
P_\sigma = \sigma_1 C_1'C_1 + \sigma_2 C_2'C_2 + \sum_{j=3}^{n} \sigma_j v_jv_j'
\end{equation*}
for any arbitrary combination of the eigenvalues $\sigma_1,\sigma_2,\sigma_j>0$.  We now evaluate the Riccati-like operator on an arbitrary $P_\sigma\in \mathcal{X}_\sigma$ and for an arbitrary action $u$
\begin{align*}
g(P_\sigma,u) 
\!=& A\left(P_\sigma^{-1} \!+\! \gamma_1C'_1R_1^{-1}C_1 \!+\! \gamma_2C'_2R_2^{-1}C_2 \right)^{\!-1}\!A' \!+\! Q \\
=& \!\begin{multlined}[t]
A\Big(\sigma_1^{-1} C_1'C_1 + \sigma_2^{-1} C_2'C_2 + \sum\nolimits_{j} \sigma_j^{-1} v_j'v_j \\  + \gamma_1C'_1R_1^{-1}C_1 + \gamma_2C'_2R_2^{-1}C_2 \Big)^{-1}A' + Q 
\end{multlined}\\
=& \!\begin{multlined}[t]
A\left(\frac{R_1\sigma_1}{R_1 + \gamma_1\sigma_1}  C_1'C_1 + \frac{R_2\sigma_2}{R_1 + \gamma_2\sigma_2}  C_2'C_2\right.\\ \left. + \sum\nolimits_{j} \sigma_j v_j'v_j \right)A' + Q 
\end{multlined}
\end{align*}
where the dependence of $\gamma$ on $u$ is omitted for clarity. 
Based on this,  we get
\begin{multline*}
\mathcal{C}(P_\sigma,(1,0)) - \mathcal{C}(P_\sigma,(0,1))= \\
q_{1}\frac{\sigma_2^2}{\sigma_2 \!+\! R_2}\mathrm{Tr}(AC_2'C_2A') - q_{2}\frac{\sigma_1^2}{\sigma_1 \!+\! R_1}\mathrm{Tr}(AC_1'C_1A')
\end{multline*}
Within the set $\mathcal{X}_\sigma$ we now restrict to the subset of matrices $P_{\bar\sigma}$ characterized by $\bar\sigma_1,\bar\sigma_2,\bar\sigma_j$ for which $\mathcal{C}(P_{\bar\sigma},(1,0)) = \mathcal{C}(P_{\bar\sigma},(0,1))$, namely satisfying
\begin{equation*}
\frac{\bar\sigma_2^2}{\bar\sigma_2 \!+\! R_2}\mathrm{Tr}(AC_2'C_2A') = \frac{q_{1}}{q_{2}}\frac{\bar\sigma_1^2}{\bar\sigma_1 \!+\! R_1}\mathrm{Tr}(AC_1'C_1A')
\end{equation*}
Now consider
\begin{align*}
\mathcal{C}(&P_{\bar\sigma},(1,1)) - \mathcal{C}(P_{\bar\sigma},(1,0)) \\
&=\mathcal{C}(P_{\bar\sigma},(1,1)) - \frac{1}{2}(\mathcal{C}(P_{\bar\sigma},(1,0))+ \mathcal{C}(P_{\bar\sigma},(1,0))) \\
&=\!\begin{multlined}[t]
\left(\frac{q_{1}}{2} - (p_{11} \!+\! p_{10})\right) \frac{\bar\sigma_1^2}{\bar\sigma_1 \!+\! R_1} \mathrm{Tr}(AC_1'C_1A') \\
+ \left(\frac{q_{2}}{2} - (p_{11} \!+\! p_{01})\right) \frac{\bar\sigma_2^2}{\bar\sigma_2 \!+\! R_2} \mathrm{Tr}(AC_2'C_2A') + \mu 
\end{multlined} \\
&=\!\begin{multlined}[t]
\left(q_{1} - (p_{11} \!+\! p_{10}) - (p_{11} \!+\! p_{01})\frac{q_{1}}{q_{2}}\right)\frac{\bar\sigma_1^2}{\bar\sigma_1\!+\! R_1}\cdot\\ \cdot\mathrm{Tr}(AC_1'C_1A')  + \mu 
\end{multlined}
\end{align*}
Since the term between parentheses is negative, we can see that the right hand side is monotonically decreasing with respect to $\bar{\sigma}_1$ and unbounded. It follows that $\exists \bar{\bar{\sigma}}_1 \!>\! 0$ such that $\mathcal{C}(P_{\bar\sigma},(1,1)) < \mathcal{C}(P_{\bar\sigma},(1,0))$ for $\bar\sigma_1>\bar{\bar{\sigma}}_1$. We can conclude that the action $u=(1,1)$ is optimal at least that for the matrices $P_{\bar{\sigma}}$ with $\bar\sigma_1>\bar{\bar{\sigma}}_1$, $\bar\sigma_2$ such that  $\mathcal{C}(P_{\bar\sigma},(1,0)) = \mathcal{C}(P_{\bar\sigma},(0,1))$, and arbitrary $\bar\sigma_j>0$. 
\end{pf*}\vspace*{-15pt}

\section{Proofs of Sec. \ref{sec:decsys}}
For sake of simplicity we consider the easiest case with two sensors.
We use the trick of splitting the cost in two parts, based on the following decomposition
\begin{align*}
\big(&P^{-1} + \gamma_1(u)C_1'R_1^{-1}C_1 
+ \gamma_2(u)C_2'R_2^{-1}C_2\big)^{-1}\\
&=\!\left[\!\begin{array}{cc}
P_1^{-1} \!+\! \gamma_1(u)C_{10}'R_1^{-1}C_{10} & 0 \\
0 & P_1^{-1} \!+\! \gamma_2(u)C_{20}'R_2^{-1}C_{20}
\end{array}\!\right]^{\!-1}
\end{align*}
\begin{pf*}{\textbf{PROOF of Proposition \ref{th:decsys}}}
We can express
\begin{align*}
\mathbb{E}&[\mathrm{Tr}(\!A(\!P^{-1} \!+\! \gamma_1(u)C_1'R_1^{-1}C_1 \!+\! \gamma_2(u)C_2'R_2^{-1}C_2\!)^{\!-1}\!A')|P,u]\\
&=\begin{multlined}[t]
\mathbb{E}[\mathrm{Tr}(A_1(P_1^{-1} + \gamma_1(u)C_{10}'R_1^{-1}C_{10})^{-1}A'_1|P,u]+ \\ 
\mathbb{E}[\mathrm{Tr}(A_2(P_2^{-1} + \gamma_2(u)C_{20}'R_2^{-1}C_{20})^{-1}A'_2)|P,u]
\end{multlined}\\
&=\begin{multlined}[t]
(1\!-\!p_1(u))\mathrm{Tr}(A_1P_1A'_1) \!+\! (1\!-\!p_2(u))\mathrm{Tr}(A_2P_2A'_2) \\ 
+p_1(u)\mathrm{Tr}(A_1(P_1^{-1} \!+\! C_{10}'R_1^{-1}C_{10})^{-1}A'_1) \vspace*{-2pt} \\ 
+p_2(u)\mathrm{Tr}(A_2(P_2^{-1} \!+\! C_{20}'R_2^{-1}C_{20})^{-1}A'_2)
\end{multlined}
\end{align*}
so that we can rewrite the one-step-ahead cost as
\begin{align*}
\mathcal{C}(P,u)\!=\begin{multlined}[t]
\!(1\!-\!p_1(u)\!)\mathrm{Tr}(A_1P_1A'_1) \!+\! (1\!-\!p_2(u)\!)\mathrm{Tr}(A_2P_2A'_2)\\
+p_1(u)\mathrm{Tr}(A_1(P_1^{-1} \!+\! C_{10}'R_1^{-1}C_{10})^{-1}A'_1) \vspace*{-2pt} \\ 
+p_2(u)\mathrm{Tr}(A_2(P_2^{-1} \!+\! C_{20}'R_2^{-1}C_{20})^{-1}A'_2) \vspace*{-2pt} \\ 
+Q_1 + Q_2+ \bar{\mu} u \hspace*{10pt} 
\end{multlined}
\end{align*}
Thus, by defining 
\begin{align*}
\psi_i(P_i)
&=\mathrm{Tr}(A_iP_iA'_i) - \mathrm{Tr}(A_i(P_i^{-1} \!+\! C_{i0}'R_i^{-1}C_{i0})^{-1}A'_i)\\
&=\mathrm{Tr}(A_iP_iC'_{i0}(C_{i0}P_iC_{i0}'+ R_i)^{-1}C_{i0}P_iA'_i) \geq 0
\end{align*}
where we used the Matrix Inversion Lemma, the minimization of $J(P,u)$ is equivalent to the maximization of $\psi_1(P_1)p_1(u)+\psi_2(P_2)p_2(u)+ \bar{\mu} u$, concluding the proof.
\end{pf*}
\begin{lem}
The function $\psi_i(X)$ is monotonically increasing with respect to  $X$.	
\end{lem}
\begin{pf}
Define $\overline{\psi}_i(X)=\mathrm{Tr}(X-(X^{-1}+C'_iR_i^{-1}C_i)^{-1})$. Note that if $\overline{\psi}_i(X)$ is monotonically increasing, the same holds for $\psi_i(X)$.
Given $f(\cdot)$ a differentiable function of an (invertible) matrix $X^{-1}$, from \cite{he2007convolutive} (see also Sec. 2.2 of \cite{petersen2012matrixcookbook}) we have that 
\begin{equation*}
\dfrac{df(X)}{dX^{-1}}=-X'\dfrac{df(X)}{dX}X'
\end{equation*}
Let $f(X)=\mathrm{Tr}((X^{-1}+C'_{i0}R_i^{-1}C_{i0})^{-1}$. For $X$ symmetric, it follows that
\begin{align*}
\dfrac{df(X)}{dX} 
&=\dfrac{d\mathrm{Tr}((X^{-1}+C'_{i0}R_i^{-1}C_{i0})^{-1})}{dX}\\
&=-X^{-1}\dfrac{d\mathrm{Tr}((X^{-1}+C'_{i0}R_i^{-1}C_{i0})^{-1})}{dX^{-1}}X^{-1}\\
&=X^{-1}(X^{-1}+C'_{i0}R_i^{-1}C_{i0})^{-2}X^{-1}
\end{align*}
where we use Equation (64) from \cite{petersen2012matrixcookbook}. Then we have
\begin{align*}
\dfrac{d\overline{\psi}_i(X)}{dX}
&= I - X^{-1}(X^{-1}+C'_{i0}R_i^{-1}C_{i0})^{-2}X^{-1}
\end{align*}
Pre and post-multiplying by $X$ we obtain $X^2 - (X^{-1}+C'_{i0}R_i^{-1}C_{i0})^{-2} \geq 0$ concluding the proof.
\end{pf}
\begin{pf*}{\textbf{PROOF of Proposition \ref{th:decsysmonotone}}}
Consider $P_2$ and $[u]_2$ fixed. We highlight this by denoting $\mathcal{C}(P_1,[u]_1)=\mathcal{C}(P,u)$. Moreover let 
$$\bar{P}=\left[ \begin{array}{cc}
\bar{P}_1 & 0 \\ 0 & P_2
\end{array}\right]$$
with $\bar{P}_1\geq P_1$ and $\bar{u}=([\bar{u}]_1,[u]_2)$ with $[\bar{u}]_1\geq [u]_1$.
Then 
\begin{align*}
\mathcal{C}&(\bar{P}_1,[u]_1) \!-\! \mathcal{C}(P_1,[u]_1) \\
&=\begin{multlined}[t]
(1\!-\!p_1(u))\mathrm{Tr}(A_1\bar{P}_1A'_1) + (1\!-\!p_1(u))\mathrm{Tr}(A_1P_1A'_1)\\
+p_1(u)\mathrm{Tr}(A_1(\bar{P}_1^{-1} + C_{10}'R_1^{-1}C_{10})^{-1}A'_1) \hspace*{-25pt}\vspace*{-2pt} \\ 
+p_1(u)\mathrm{Tr}(A_1(P_1^{-1} + C_{10}'R_1^{-1}C_{10})^{-1}A'_1)
\end{multlined}\\
&=\mathrm{Tr}(A_1(\bar{P}_1\!-\!P_1)A'_1) - p_1(u)(\psi_1(\bar{P}_1) \!-\! \psi_1(P_1))\\
&\geq \mathrm{Tr}(A_1(\bar{P}_1\!-\!P_1)A'_1) - p_1(\bar{u})(\psi_1(\bar{P}_1) \!-\! \psi_1(P_1))\\
&=\mathcal{C}(\bar{P}_1,[\bar{u}]_1) \!-\! \mathcal{C}(P_1,[\bar{u}]_1) 
\end{align*}
where the inequality holds because $\psi_i(P_i)$ is monotonically increasing w.r.t. $P_i$ and $p_i(u)$ is monotonically increasing w.r.t. $[u]_i$ for fixed $[u]_j$, $j\neq i$.
We conclude that
\begin{equation*}\vspace*{-2pt}
\mathcal{C}(\bar{P}_1,[\bar{u}]_1) - \mathcal{C}(P_1,[\bar{u}]_1) \leq \mathcal{C}(\bar{P}_1,[u]_1) - \mathcal{C}(P_1,[u]_1)\vspace*{-2pt}
\end{equation*}
From Theorem 2.6.1 by \cite{topkis1998supermodularity} $\mathcal{C}(P_1,[u]_1)$ is submodular in $(P_1,[u]_1)$.
From Theorem 2.8.1 by \cite{topkis1998supermodularity} submodularity is a sufficient condition for optimality of monotone increasing policies. See also the proof of Theorem 6.1 in \cite{nourian2014optimal}. In particular since $\mathcal{C}(P_1,[u]_1)$ is submodular in $(P_1,[u]_1)$, then $u^*_1(P_1) \!=\! \arg\min \mathcal{C}(P_1,[u]_1)$ is non-decreasing in $P_1$.
\end{pf*}\vspace{-15pt}

\bibliographystyle{plain}
\bibliography{bibfull,references} 

\end{document}